\documentclass[twocolumn,aps,pra]{revtex4-2}
\usepackage{amssymb}
\usepackage{amsfonts}
\usepackage{amsmath}
\usepackage{graphicx}

\input{tcilatex}

\begin{document}

\title{Optimized discrete Wigner representations and non-stabilizerness in
qubit systems}
\author{A.B. Klimov, C. Mu\~{n}oz, I. Sainz}
\affiliation{ Departamento de F\'isica, Centro Universitario de Ciencias Exactas e Ingenier\'ia, Universidad de Guadalajara,
Revoluci\'on 1500, 44410 Guadalajara, M\'exico}
\date{\today }

\begin{abstract}
The apparent absence of a stabilizer-specific phase-space description for
qubit systems is not fundamental, but arises from fixing a state-independent
discrete Wigner map. Allowing the phase-space representation to adapt to the
state through a simple optimization procedure, we prove that every qubit
stabilizer state admits an exact delta-function discrete Wigner
representation for any fixed phase-space partition. This optimization can be
implemented directly at the level of measurement statistics by selecting
maximal probabilities across mutually unbiased bases forming the partition.

This construction naturally leads to a non-stabilizer witness defined by the
optimized value of the Wigner function at the origin. Stabilizer states
saturate the universal lower bound  of this witness for pure states,
whereas a value strictly larger than the bound certifies non-stabilizerness.
Although the witness is not a finite-size resource monotone, it becomes
sharply concentrated and effectively invariant under random Clifford
transformations in the large-qubit limit, suggesting an asymptotically
linear growth law for broad families of non-stabilizer states. Our
results restore a meaningful geometric phase-space characterization of qubit
stabilizer structure and provide an operational diagnostic of
non-stabilizerness.


\end{abstract}

\maketitle

\section{Introduction}

Although entanglement has long been viewed as the central ingredient of
quantum technologies \cite{bennett,ekert,supremacy}, entanglement alone does
not account for quantum computational advantage. A paradigmatic example is
stabilizer fault-tolerant quantum computing: states like
Greenberger-Horne-Zeilinger (GHZ), despite being highly entangled, can be
efficiently simulated classically according to the Gottesman-Knill theorem
\cite{gottesman,aaronson,nest2010}.  To surpass classical simulation
capabilities \cite{bravyi,bravyi2,howard1,beverland}, a more refined
nonclassical resource, known as magic \cite{bravyi}, which can be injected
through non-stabilizer states, is essential.

In systems of power of odd prime-power dimension $p^{n}$, the deviation
from the stabilizer formalism admits a particularly transparent
phase-space description through the discrete Wigner function (DWF) $W(u)$, $
u\in \mathbb{F}_{p^{n}}\times \mathbb{F}_{p^{n}}$, \cite{wootters87,
gibbons,galvao,bjorkDWF}: a pure state has a nonnegative Wigner function if
and only if it is a stabilizer state \cite{gross}, and only Clifford
operations preserve nonnegativity \cite{gross,zhu16}. As a result,
negativity of the DWF provides a direct and physically meaningful signature
of magic \cite{mari,veitch12,resourceStab,koukou}. In this framework, magic,
Wigner-function negativity, and contextuality become equivalent notions \cite{contextqd1,contextqd2,contextqd3}, and the discrete Wigner function
provides the unique noncontextual ontological model compatible with
stabilizer subtheories \cite{contextqd3}.

For qubits, however, this phase-space picture faces fundamental limitations:
there is no discrete Wigner function simultaneously satisfying covariance
under the Clifford transformations and nonnegativity for stabilizer states
\cite{raussendorf17,zhu16,zhu17,cohomology} (technically, no factorized
Wigner map beyond 2-qubit systems meets the tomographic requirement).
Consequently, no noncontextual, state-independent, ontological model exists
for general multi-qubit stabilizer subtheories \cite{zhu16,zhu17,cohomology,catani,raussendorf17,catani18,lillystone,rebits},
and the physical and geometric role of the discrete Wigner function in qubit
systems, as well as its relation to non-stabilizerness, is still uncertain.
As a result, most existing approaches to qubit magic rely on algebraic
methods rather than phase-space geometry \cite{beverland,sre,espectrum}.

A central technical obstacle lies in the construction of the $n$-qubit
discrete Wigner function itself. The DWF depends not only on the choice of a
phase-space partition associated with a complete set of mutually unbiased
bases (MUBs) \cite{MUB,wootters87,cormick,bjork2007,curvas}, but also on a
vast gauge freedom in the choice of cocycle phases determining the sets of
stabilizer operators. For any fixed partition, this freedom gives rise to $
\left( 2^{n}\right) ^{2^{n}+1}$ (mostly) inequivalent phase-space
arrangements within any fixed partition, corresponding to different choices
of phase conventions consistent with the same underlying stabilizer
structure \cite{gibbons,curvas1,fasesqb,faseqb1,faseRauss,ernesto}.

Physically relevant properties of the discrete Wigner function, including
positivity and localization for stabilizer states, depend sensitively on the
chosen arrangement \cite{catani,raussendorf17,ernesto,faseRauss} (within a
fixed partition). In particular, a stabilizer state is not necessarily
represented as a nonnegative Wigner distribution for a generic choice of
phase conventions \cite{cormick,ernesto}. In addition, stabilizer states
related by Clifford transformations can exhibit markedly different
Wigner-function behavior within a fixed representation \cite{cormick,Munoz17}.
Thus, in contrast to the odd prime-power dimensional case, the positivity
of the discrete Wigner function is not an intrinsic characteristic of qubit
stabilizer states, and the notion of classicality inherently depends on the
phase-space structure.

In this work, we show that this apparent drawback reflects not the absence
of stabilizer structure in qubit phase space, but rather the restriction to
state-independent phase-space representations. With a simple state-adapted
optimization, we prove that every $n$-qubit stabilizer state admits the same
delta-function form of the Wigner representation for any fixed phase-space
partition. The existence of such state-adapted representations is guaranteed
by the cocycle (gauge) freedom inherent in the qubit discrete phase-space
\cite{raussendorf17,faseRauss,ernesto,park}. In addition, the optimization
itself can be implemented entirely in terms of measurement probabilities
across mutually unbiased bases, without explicitly determining cocycle
phases. As a result, for pure states the optimized value of the Wigner
function at the origin satisfies the sharp lower bound $W^{opt}(0)\geq 1$,
which is saturated by stabilizer states. This establishes a structural
analogue of Hudson's theorem for qubit systems and restores a
stabilizer-specific geometric interpretation of the discrete Wigner
function. On the other hand, saturation, $W^{opt}(0)=1$, leads to the
so-called flat states, i.e., states whose nonzero probabilities are constant
on their supports simultaneously in all MUBs used for the phase-space
construction. We prove that flatness of a state implies its stabilizerness
for one and two qubits (the same analysis extends directly to three qubits),
and for a specific class of $n$-qubit states. We also provide heuristic
and exhaustive numerical evidence supporting the conjectured equivalence of
flatness and stabilizerness, or equivalently, the additivity of the centered
supports of all-MUB-flat states.

In this framework, we introduce a non-stabilizer witness based on the
optimized value of the discrete Wigner function at the origin, which admits
a simple interpretation in terms of the maximal measurement probabilities
across mutually unbiased bases used to construct the phase-space partition.
This witness involves an explicit state-dependent optimization and is
therefore not a resource monotone at finite system size. Nevertheless, it
exhibits several notable asymptotic properties. For Haar-random pure states,
the witness can be computed analytically, and its expectation is independent
of the chosen MUB partition, while its variance vanishes as the number of
qubits increases, providing a benchmark for non-stabilizerness in $n$%
-qubit systems. Independently, for a fixed sequence of pure states, the
witness variance under uniformly sampled Clifford transformations also
vanishes, so the witness concentrates around its state-dependent
Clifford-orbit mean. We prove linear-order growth for a class of sparse
states; both numerical evidence and heuristic arguments are provided to
support the same behavior for broader state families, thus offering a
physically transparent diagnostic of non-stabilizerness in macroscopic qubit
systems.

The paper is organized as follows: In Sec.II, we briefly review the
stabilizer formalism and its realization in discrete phase-space,
emphasizing the role of cocycle phases and arrangements. In Sec. III, we
introduce the optimization procedure and prove the structural theorem
characterizing stabilizer states via delta-shaped Wigner functions. In Sec.
IV, we introduce the non-stabilizer witness and analyze its properties,
asymptotic behavior, and relation to existing measures of magic. Finally,
Sec.V summarizes the main conclusions and their scope.

\section{Stabilizer states in the discrete phase-space}

\subsection{Stabilizers and stabilizer states}

\textit{Pauli group and displacement operators}

It is convenient to describe the geometric properties of stabilizer states
of the $n$-qubit systems by adopting the finite field notation, commonly
employed in the discrete phase-space approach \cite{wootters87,gibbons,galvao,veitch12,rebits,curvas,vourdas2004,paz2005}.
An orthonormal (computational) basis $|k_{1},\ldots ,k_{n}\rangle $ ($k_{i}=0,1$) in $\mathcal{H}_{2^{n}}=\mathcal{H}_{2}^{\otimes n}$ is identified with
elements of the finite field $\mathbb{F}_{2^{n}}$, $|\mathbb{F}
_{2^{n}}|=2^{n}=d$, according to $\kappa =\sum_{i=1}^{n}k_{i}\,\theta _{i}$
\begin{equation}
|k_{1},\ldots ,k_{n}\rangle \Leftrightarrow |\kappa \rangle ,~\langle \kappa
^{\prime }|\kappa \rangle =\delta _{\kappa ^{\prime },\kappa },
\label{basis}
\end{equation}
where $\{\theta _{i}\}_{i=1}^{n}$ is a self-dual basis of $\mathbb{F}_{2^{n}}
$ satisfying $tr(\theta _{i}\,\theta _{j})=\delta _{ij}$, $tr(\alpha
)=\alpha +\alpha ^{2}+\ldots +\alpha ^{2^{n-1}}$ \cite{FF}. Throughout, $tr$
denotes the finite-field trace, whereas $Tr$ denotes the Hilbert-space
operator trace. This identification allows one to treat states and operators
as functions on $\mathbb{F}_{2^{n}}$, simplifying both algebraic operations
and the geometric interpretation.

The Pauli group $\mathcal{P}_{n}$ is generated (up to overall phases $\pm
1,\pm i$) by $d^{2}$ monomials $\{\hat{Z}_{\alpha }\hat{X}_{\beta }$, $
\alpha ,\beta \in \mathbb{F}_{2^{n}}\}$, where
\begin{eqnarray*}
\hat{Z}_{\alpha } &=&\sum_{\kappa }\chi (\alpha \kappa )\left\vert \kappa
\right\rangle \langle \kappa |,\quad \hat{X}_{\beta }=\sum_{\kappa
}\left\vert \kappa +\beta \right\rangle \langle \kappa |, \\
\hat{Z}_{\alpha }\hat{X}_{\beta } &=&\chi (\alpha \beta )\hat{X}_{\beta }
\hat{Z}_{\alpha },\quad \chi (\alpha )=(-1)^{tr(\alpha )};
\end{eqnarray*}
here $\chi (\alpha )$ are additive characters on $\mathbb{F}_{2^{n}}$, $\chi
(\alpha +\beta )=\chi (\alpha )\chi (\beta )$ and $tr(\alpha \beta
)=\sum_{i=1}^{n}a_{i}b_{i}\in \mathbb{Z}_{2}$.

The commutation relation of unitary Hermitian displacement operators
\begin{equation}
\hat{D}(\alpha ,\beta )=\Phi _{\alpha ,\beta }\hat{Z}_{\alpha }\hat{X}
_{\beta }\equiv \hat{D}_{u},\;u=(\alpha ,\beta ),  \label{Dab}
\end{equation}
where $\Phi _{\alpha ,\beta }=\Phi _{\alpha ,\beta }^{\ast }\chi (\alpha
\beta )$ and $\Phi _{0,0}=1$, is determined by the symplectic form
\begin{equation}
\lbrack u,v]=[(\alpha ,\beta ),(\alpha ^{\prime },\beta ^{\prime
})]=tr\left( \alpha \beta ^{\prime }+\alpha ^{\prime }\beta \right) ,
\label{uv}
\end{equation}
 according to
\begin{equation}
\hat{D}_{u}\hat{D}_{v}=(-1)^{[u,v]}\hat{D}_{v}\hat{D}_{u}.  \label{DDu}
\end{equation}

A stabilizer set $\lambda $  of $2^{n}$ commuting displacement operators,
\begin{equation}
\hat{D}^{\lambda }\left( \tau \right) =\hat{V}_{\lambda }\hat{Z}_{\tau }\hat{
V}_{\lambda }^{\dagger }=\hat{D}^{\lambda \dagger }\left( \tau \right) ,
\label{Dg}
\end{equation}
is obtained by conjugating the set $\{\hat{Z}_{\tau },\tau \in \mathbb{F}
_{2^{n}}\}$ by a Clifford unitary $\hat{V}_{\lambda }$. The corresponding
stabilizer states $\{|\psi _{\kappa }^{\lambda }\rangle =\hat{V}_{\lambda
}|\kappa \rangle ,$ $\kappa \in \mathbb{F}_{2^{n}}\}$ form a basis in $
\mathcal{H}_{2^{n}}$,
\begin{equation}
\hat{D}^{\lambda }\left( \tau \right) |\psi _{\kappa }^{\lambda }\rangle
=\chi (\kappa \tau )|\psi _{\kappa }^{\lambda }\rangle ,\;\sum_{\kappa
}|\psi _{\kappa }^{\lambda }\rangle \langle \psi _{\kappa }^{\lambda }|=\hat{%
I}.  \label{SS1}
\end{equation}
Throughout, we will refer to all eigenstates of the stabilizer set (\ref{Dg}) (and not only to that of unity eigenvalue, i.e., $\kappa =0$) as
stabilizer states.

The operators (\ref{Dg})\ inherit the Abelian group structure,
\begin{equation}
\hat{D}^{\lambda }\left( \tau \right) \hat{D}^{\lambda }\left( \tau ^{\prime
}\right) =\hat{D}^{\lambda }(\tau +\tau ^{\prime }),  \label{DD}
\end{equation}
and can be written as Pauli monomials (\ref{Dab})
\begin{equation}
\hat{D}^{\lambda }\left( \tau \right) =\Phi ^{\lambda }\left( \tau \right)
\hat{Z}_{\alpha _{\lambda }(\tau )}\hat{X}_{\beta _{\lambda }(\tau )}\equiv
\hat{D}_{\gamma _{\lambda }(\tau )},  \label{DZX}
\end{equation}
for a stabilizer $\Gamma ^{\lambda }$-set:
\begin{equation}
\Gamma ^{\lambda }=\{\gamma _{\lambda }(\tau )=\left( \alpha _{\lambda
}(\tau ),\beta _{\lambda }(\tau )\right) ,\;\tau \in \mathbb{F}
_{2^{n}}\}\subset V,  \label{G}
\end{equation}
where $ V=\mathbb{F}_{2^{n}}\times \mathbb{F}_{2^{n}}$, and $\alpha _{\lambda }(\tau ),\beta _{\lambda }(\tau )$ are additive
functions \cite{planar f},
\begin{eqnarray*}
\alpha _{\lambda }(\tau +\tau ^{\prime }) &=&\alpha _{\lambda }(\tau
)+\alpha _{\lambda }(\tau ^{\prime }), \\
\beta _{\lambda }(\tau +\tau ^{\prime }) &=&\beta _{\lambda }(\tau )+\beta
_{\lambda }(\tau ^{\prime }).
\end{eqnarray*}
The phases $\Phi _{\alpha ,\beta }\left( \tau \right) $ satisfy the cocycle
condition \cite{faseRauss,fasesqb},
\begin{equation}
\Phi ^{\lambda }\left( \tau \right) \Phi ^{\lambda }\left( \tau ^{\prime
}\right) =\Phi ^{\lambda }\left( \tau +\tau ^{\prime }\right) \chi \left(
\alpha _{\lambda }(\tau ^{\prime })\beta _{\lambda }(\tau )\right) ,
\label{Phi}
\end{equation}
which is a direct consequence of the group law (\ref{DD}). This implies the
commutativity constraint:
\begin{equation}
tr\left( \alpha _{\lambda }(\tau ^{\prime })\beta _{\lambda }(\tau )\right)
=tr\left( \alpha _{\lambda }(\tau )\beta _{\lambda }(\tau ^{\prime })\right)
,  \label{CC}
\end{equation}%
on the elements of the set (\ref{G}), defining the maximal Abelian subgroup $
\{\hat{Z}_{\alpha _{\lambda }(\tau )}\hat{X}_{\beta _{\lambda }(\tau )},\tau
\in \mathbb{F}_{2^{n}}\}$ according to (\ref{uv})-(\ref{DDu}). The functions
$\alpha _{\lambda }(\tau ),\beta _{\lambda }(\tau )$ are not necessarily
invertible \cite{curvas}, however, the number of different ordered pairs $
\left( \alpha _{\lambda }(\tau ),\beta _{\lambda }(\tau )\right) $ is $2^{n}$.

A displacement operator $\hat{D}_{u}$ not {belonging } to a particular
stabilizer set $\{\hat{D}_{\gamma _{\lambda }(t)}\}$ acts on that stabilizer
basis $\{|\psi _{\kappa }^{\lambda }\rangle \}$ through the index shifts:
\begin{align*}
\hat{D}_{\gamma _{\lambda }(\tau )}\hat{D}_{u}|\psi _{\kappa }^{\lambda
}\rangle & =(-1)^{[u,\gamma _{\lambda }(\tau )]}\chi (\kappa \tau )\hat{D}
_{u}|\psi _{\kappa }^{\lambda }\rangle \\
& =\chi \!\left[ (\kappa +\sigma _{\lambda }(u))\tau \right] \hat{D}
_{u}|\psi _{\kappa }^{\lambda }\rangle ,
\end{align*}
that is
\begin{equation}
\hat{D}_{u}|\psi _{\kappa }^{\lambda }\rangle =e^{i\vartheta _{\lambda
}(u,\kappa )}|\psi _{\kappa +\sigma _{\lambda }(u)}^{\lambda }\rangle ,
\label{label shift}
\end{equation}
where the induced label change is the element $\sigma _{\lambda }(u)\in
\mathbb{F}_{2^{n}}$ defined by
\begin{equation}
\func{tr}\!\left( \sigma _{\lambda }(u)\tau \right) =[u,\gamma _{\lambda
}(\tau )],  \label{sigma}
\end{equation}
for every $\tau \in \mathbb{F}_{2^{n}}$. Note that $\sigma _{\lambda }(u)$
is unique since the trace pairing is nondegenerate. The arbitrary eigenvalue
characters (used below) are equivalent to the conventional +1-eigenvalue
definition after an appropriate rephasing of the commuting Pauli generators.

\textit{Cocycle phases and gauge freedom}

It follows from (\ref{Phi}) that
\begin{equation}
\left( \Phi ^{\lambda }\left( \tau \right) \right) ^{2}=\chi \left( \alpha
_{\lambda }(\tau )\beta _{\lambda }(\tau )\right) ,  \label{SE}
\end{equation}
so that the cocycle $\Phi ^{\lambda }\left( \tau \right) $ is fixed up to a
sign \cite{fasesqb, faseqb1} at any $\left( \alpha _{\lambda }(\tau ),\beta
_{\lambda }(\tau )\right) $. Multiplication of a solution of (\ref{Phi}) by
a character
\begin{equation}
\Phi _{\eta }^{\lambda }\left( \tau \right) \equiv \chi (\tau \eta _{\lambda
})\Phi ^{\lambda }\left( \tau \right) ,  \label{Phinu}
\end{equation}%
generates an equivalent cocycle, corresponding to a relabelling of
stabilizer eigenstates, $|\psi _{\kappa }^{\lambda }\rangle \rightarrow
|\psi _{\kappa +\eta _{\lambda }}^{\lambda }\rangle $, and reflecting a
\textquotedblleft gauge\textquotedblright\ freedom in the cocycle choice.
Explicit phase-constrained expansions of stabilizer states in Pauli
eigenbases are given in Appendix A.

\subsection{Stabilizer curves in the discrete phase space}

The discrete phase-space (DPS) of an $n$-qubit system is a $2^{n}\times
2^{n} $ grid endowed with a finite geometry \cite{gibbons,galvao,cormick,curvas,fasesqb,vourdas2004,paz2005}. Phase-space
points $u=(\alpha ,\beta )\in V=\mathbb{F}_{2^{n}}$ $\times \mathbb{F}
_{2^{n}}$, label Pauli monomials $\hat{Z}_{\alpha }\hat{X}_{\beta }$, where
the coordinate axes correspond to commuting families $\{\hat{Z}_{\alpha }\}$
and $\{\hat{X}_{\beta }\}$.

\medskip
\textit{Stabilizer curves and commuting subgroups}

A stabilizer set (\ref{G}) defines an isotropic subspace (a stabilizer
curve) \cite{curvas,curvas1} (\ref{G}) containing $2^{n}$ distinct points.
The curve $\Gamma ^{\lambda }$ is a maximal isotropic additive subgroup
(Lagrangian subspace) of $V$, isomorphic to $(Z_{2})^{n}$: $\gamma _{\lambda
}(\tau )+\gamma _{\lambda }(\tau ^{\prime })=\gamma _{\lambda }(\tau +\tau
^{\prime })$. The associated displacement operators additionally carry
cocycle phases $\Phi _{\alpha ,\beta }^{\lambda }\left( \tau \right) $
assigned to the points of $\Gamma ^{\lambda }$. All stabilizer curves pass
through the origin, corresponding to the identity operator in the stabilizer
sets $\{\hat{D}_{\gamma _{\lambda }(\tau )}\}$.

The parameter $\tau $ is the coordinate along the curve $\Gamma ^{\lambda }$, and $\gamma _{\lambda }(\tau )$ can be considered as an additive map
$\gamma _{\lambda }(\tau ):$ $\mathbb{F}_{2^{n}}\rightarrow V$. The map (\ref{sigma}), $V\rightarrow \mathbb{F}_{2^{n}}$, is an additive functional with
$\ker \sigma _{\lambda }=\Gamma ^{\lambda }$, since $\sigma _{\lambda
}(\Gamma ^{\lambda })=0$, and conversely, $[u,\gamma _{\lambda }(\tau )]=0$,
for all $\tau \in \mathbb{F}_{2^{n}}$ implying that $u\in (\Gamma ^{\lambda
})^{\perp }$, the symplectic orthogonal of $\Gamma ^{\lambda }$, and for
maximal isotropic subspaces, $(\Gamma ^{\lambda })^{\perp }=\Gamma ^{\lambda
}$.

Each stabilizer curve $\Gamma ^{\lambda }$ defines a $\lambda $-striation,
consisting of $\Gamma ^{\lambda }$ and $2^{n}-1$ affine translates, called
\textquotedblleft parallel\textquotedblright\ curves $\{\Gamma ^{\lambda }+u$, $u\in V\}$. An affine translate is not an additive subgroup and therefore
does not define a stabilizer group. Two points $u$ and $u^{\prime }$ belong
to the same translate if $u+u^{\prime }\in \Gamma ^{\lambda }$, which
implies $\sigma _{\lambda }(u)=\sigma _{\lambda }(u^{\prime })$. In this
context, $\sigma _{\lambda }(u)$ assigns a single label to all points of a
translate $\Gamma ^{\lambda }+u$, formalized as $\sigma _{\lambda}:V\rightarrow V/\Gamma ^{\lambda }\simeq \mathbb{F}_{2^{n}}$. Thus, for a
fixed stabilizer curve $\Gamma $, $\sigma _{\lambda }(\Gamma )$ represents a
projection onto the space of translates that are transverse to $\Gamma^{\lambda }$.

\medskip
\textit{Intersections and overlaps}

The crossing points of two different stabilizer curves $\Gamma ^{\lambda }$
and $\Gamma ^{\lambda ^{\prime }}$ form an isotropic subspace of $V$, so
that their intersection $\Gamma ^{\lambda }\cap \Gamma ^{\lambda ^{\prime }}$
is labeled by elements of an additive subgroup. The corresponding
basis-state overlaps are determined by this intersection together with the
compatibility of their stabilizer eigenvalues \cite{geomStab}.

Let $\Gamma \subset V$ be the maximal isotropic subgroup associated with a
stabilizer state $s_{\Gamma ,\xi }\rangle $, such that
\begin{equation}
\hat{D}_{u}|s_{\Gamma ,\xi }\rangle =\chi _{\xi }(u)|s_{\Gamma ,\xi }\rangle
,\;u\in \Gamma ,\quad \xi \in \mathbb{F}_{2^{n}},  \label{ksiu}
\end{equation}%
where $\chi _{\xi }(u)=\pm 1$ are the characters defining the corresponding
eigenvalues through $\xi $. For a given curve $\Gamma ^{\lambda }$, the
subgroup
\begin{equation}
T_{\lambda }=\gamma _{\lambda }^{-1}(\Gamma \cap \Gamma ^{\lambda }),\;\dim
_{\mathbb{F}_{2}}T_{\lambda }=r_{\lambda },~ |\Gamma \cap \Gamma
^{\lambda }|=|T_{\lambda }|=2^{r_{\lambda }},  \label{Tl}
\end{equation}%
labels $2^{r_{\lambda }}$ phase-space points of the geometric intersection
of the stabilizer curve with the curve defining the $\lambda $-basis.

\textit{Proposition 1 (stabilizer flatness).}

For every $\lambda $, there is a label $\kappa _{\lambda }^{0}$ such that
\begin{equation}
p_{\lambda \kappa }^{(\Gamma )}=|\langle \psi _{\kappa }^{\lambda
}|s_{\Gamma ,\xi }\rangle |^{2}=%
\begin{cases}
|T_{\lambda }|/d, & \kappa \in \kappa _{\lambda }^{0}+H_{\lambda }, \\[1mm]
0, & \text{otherwise},%
\end{cases}
\label{sk}
\end{equation}
where
\begin{equation}
H_{\lambda }\equiv T_{\lambda }^{\perp }=\{q:\func{tr}(q\tau )=0\text{ for
all }\tau \in T_{\lambda }\}  \label{Tort}
\end{equation}%
is the centered additive support, and $\kappa _{\lambda }^{0}$ is any
solution of the set of equations,
\begin{equation}
\chi (\kappa \tau )=\chi _{\xi }(\gamma _{\lambda }(\tau )),\qquad \tau \in
T_{\lambda }.  \label{chiz}
\end{equation}%
Geometrically, $H_{\lambda }$ is the projection
\begin{equation}
H_{\lambda }=\sigma _{\lambda }(\Gamma ),  \label{sigmaT}
\end{equation}%
so the occupied labels form the affine coset $\kappa _{\lambda}^{0}+H_{\lambda }$, with $|H_{\lambda }|=d/|T_{\lambda }|=2^{n-r_{\lambda}} $. Thus, a stabilizer state is decomposed in each $\lambda $-basis as
\begin{equation*}
|s_{\Gamma ,\xi }\rangle =|H_{\lambda }|^{-1/2}\sum_{\kappa \in H_{\lambda
}}z_{\kappa }|\psi _{\kappa +\kappa _{\lambda }^{0}}^{\lambda }\rangle
,\quad |z_{\kappa }|=1,
\end{equation*}%
so that the state $|s_{\Gamma ,\xi }\rangle $ is an eigenstate of $\hat{D}%
_{\gamma _{\lambda }(\tau )}$, for all $\tau \in T_{\lambda }$, and the
probability on each occupied label is $2^{r_{\lambda }-n}$.

The proof of Proposition 1 is given in Appendix B.

Geometrically, (\ref{sigmaT}) means that $H_{\lambda }$ is the set of
transverse labels of translates $\sigma _{\lambda }(u)$, $u\in \Gamma $ of
the $\lambda $-striation intersected with a stabilizer curve $\Gamma $:

\begin{equation}
(v+\Gamma ^{\lambda })\cap \Gamma \neq \varnothing \Leftrightarrow \sigma
_{\lambda }(v)\in T_{\lambda }^{\perp },\quad v\in V.  \label{crossG1}
\end{equation}%
Algebraically, the centered support $H_{\lambda }$ is the projection $\sigma
_{\lambda }(\Gamma )$ of $\Gamma $ onto the transverse labels of the $\lambda $-striation. In addition, according to (\ref{kerG})-(\ref{Psize})
every support label has $|\Gamma \cap \Gamma ^{\lambda }|=2^{r_{\lambda }}$
pre-images in $\Gamma $, which is the projection degeneracy.

\medskip \textit{Partitions and mutually unbiased stabilizer bases}

If two stabilizer curves intersect only at the origin, then the
corresponding stabilizer sets are disjoint (except for the identity). In
this case, their eigenbases are mutually unbiased:
\begin{equation}
|\langle \psi _{\kappa ^{\prime }}^{\lambda ^{\prime }}|\psi _{\kappa
}^{\lambda }\rangle |^{2}=\delta _{\lambda ,\lambda ^{\prime }}\delta
_{\kappa ,\kappa ^{\prime }}+2^{-n}(1-\delta _{\lambda ,\lambda ^{\prime }}).
\label{UB}
\end{equation}

A phase-space partition $\pi $ (symplectic spread) is a fixed set $\Lambda $
of $2^{n}+1$ pairwise non-crossing (except at the origin) stabilizer curves $%
\{\Gamma ^{\lambda },\,\lambda \in \Lambda \}$, $\,\Gamma ^{\lambda ^{\prime
}}\cap \Gamma ^{\lambda }=\{0\}$, $\lambda ,\lambda ^{\prime }\in \Lambda $,
whose union covers the whole DPS, i.e., all nontrivial Pauli monomials. A
canonical example is the Desarguesian (ray) partition:

\begin{eqnarray}
\Gamma ^{\lambda } &=&\{(\tau ,\lambda \tau ),\;\tau \in \mathbb{F}%
_{2^{n}}\},\quad \lambda \in \mathbb{F}_{2^{n}},  \label{DM1} \\
\Gamma ^{\infty } &=&\{(0,\tau ),\;\tau \in \mathbb{F}_{2^{n}}\},
\label{DM2}
\end{eqnarray}
or equivalently%
\begin{equation}
\{\beta =\lambda \alpha ,\;\lambda \in \mathbb{F}_{2^{n}}\}\cup \{\alpha
=0\},  \label{rays}
\end{equation}%
leading to the familiar stabilizer MUB construction%
\begin{equation}
\{|\psi _{\kappa }^{\lambda }\rangle =\hat{V}_{\lambda }|\kappa \rangle
,\;\kappa ,\lambda \in \mathbb{F}_{2^{n}}\}\cup \{|\tilde{\kappa}\rangle
\;\kappa \in \mathbb{F}_{2^{n}}\},  \label{MUBray}
\end{equation}%
where the Clifford unitaries $\hat{V}_{\lambda }$ (\ref{Dg}) may be chosen
diagonal in the $X$-eigenbasis $\{|\tilde{\tau}\rangle \}$, $\hat{X}_{\nu }|%
\tilde{\tau}\rangle =\chi (\nu \tau )|\tilde{\tau}\rangle $,
\begin{eqnarray}
\hat{V}_{\lambda } &=&\sum_{\tau \in \mathbb{F}_{2^{n}}}\phi ^{\lambda
}\left( \tau \right) |\tilde{\tau}\rangle \langle \tilde{\tau}|,
\label{Vray} \\
\phi ^{\lambda }\left( \tau +\tau ^{\prime }\right) &=&\phi ^{\lambda }(\tau
)\phi ^{\lambda }(\tau ^{\prime })\chi \left( \lambda \tau \tau ^{\prime
}\right) .  \notag
\end{eqnarray}

Phase-space partitions exist in multiple (locally inequivalent) forms \cite{curvas,lawrence,romero07}, and some of them can be related by global
Clifford transformations applied simultaneously to all stabilizer sets in
the partition (unitarily inequivalent partitions were considered in \cite{Kantor}). In Appendix C we give an explicit three-qubit example of a
partition locally inequivalent to (\ref{rays}).

Note that by Proposition 1, any stabilizer state corresponding to a curve $\Gamma $ is flat in every MUB of a partition $\pi $, and the
corresponding centered supports are additive subgroups. The (centered)
support in every MUB is the projection of the same stabilizer curve $\Gamma $, viewed in a different transverse coordinate system according to (\ref{sigmaT}). For the Desarguesian partition, these projection maps take
simple forms: $\sigma _{x}(\alpha ,\beta )=\alpha $ and $\sigma _{\lambda
}(\alpha ,\beta )=\beta +\lambda \alpha $. The corresponding supports and
projection degeneracies of a stabilizer curve are derived explicitly in
Appendix D.

We call a pure state all-MUB-flat if, in every basis of the complete
stabilizer MUB, its nonzero probabilities are equal.

\textit{Proposition 2 (additive-support criterion).}

Within the class of all-MUB-flat pure states, a state is a stabilizer state
if and only if all of its centered supports are additive subgroups.

\textit{Proof.} According to Proposition 1, a stabilizer state is flat in
every basis of a phase-space partition $\pi $. The points of a stabilizer
curve $\Gamma $are distributed over all $\Gamma ^{\lambda },\lambda \in
\Lambda ,$ of $\pi $, so that,
\begin{equation*}
d-1=\sum_{\lambda \in \Lambda }\left( |\Gamma \cap \Gamma ^{\lambda
}|-1\right) =\sum_{\lambda \in \Lambda }(2^{r_{\lambda }}-1),
\end{equation*}%
and since the support size is $|H_{\lambda }|=d/2^{r_{\lambda }}$, one
obtains
\begin{equation}
\sum_{\lambda \in \Lambda }|H_{\lambda }|^{-1}=d^{-1}\sum_{\lambda \in
\Lambda }2^{r_{\lambda }}=2.  \label{sumH-1}
\end{equation}%
Conversely, let $|\psi \rangle $ be a pure state whose probability
distribution is flat on an additive centered support $H_{\lambda }$ in every
basis of the complete stabilizer MUB. Then, according to Eq. (\ref{SS1}) and
character orthogonality, one has

\begin{equation}
|\langle \psi |\hat{D}_{\gamma _{\lambda }(\tau )}|\psi \rangle |^{2}=\frac{1}{|H_{\lambda }|^{2}}|\sum_{\kappa \in H_{\lambda }}\chi (\kappa \tau )|^{2}=
\mathbf{1}_{H_{\lambda }^{\perp }}(\tau ),  \label{Davf}
\end{equation}
so that, in each striation the displacements are labeled by $\gamma
_{\lambda }(H_{\lambda }^{\perp })$. Thus, all displacement operators
labelled by
\begin{equation}
\Gamma _{\psi }=\{0\}\cup \bigcup_{\lambda \in \Lambda }\gamma _{\lambda
}(H_{\lambda }^{\perp }\setminus \{0\}).  \label{derUnionL}
\end{equation}
Since the expectation value in Eq.~(\ref{Davf}) has unit modulus for these
displacements, $|\psi \rangle $ is their common eigenvector. Pauli
displacements share this eigenvector, and thus commute pairwise. Their
labels therefore belong to a common isotropic subspace.

Since the displacements form an orthogonal operator basis, the purity
condition and Eq.(\ref{Davf}) give
\begin{equation}
d=\sum_{u\in V}|\langle \psi |\hat{D}_{u}|\psi \rangle |^{2}=1+\sum_{\lambda
\in \Lambda }\left( |H_{\lambda }^{\perp }|-1\right) =|\Gamma _{\psi }|.
\label{avD}
\end{equation}
An isotropic subspace of the $2n$-dimensional binary symplectic space
contains at most $d=2^{n}$ points. Since $\Gamma _{\psi }$ already contains $d$ distinct points, its isotropic span must coincide with $\Gamma _{\psi }$.
Therefore, $\Gamma _{\psi }$ is maximal isotropic, and $|\psi \rangle $ is a
stabilizer state. Thus, for the class of complete-MUB-flat pure states, the
additivity of all centered supports is equivalent to stabilizerness.

\textit{Phase-space arrangements and cocycle gauge }

For a fixed partition $\pi =\{\Gamma ^{\lambda }\}$, the cocycle phases
along each curve are not unique. Using the gauge freedom (\ref{Phinu}), one
may define a family of stabilizer operators on the same curve,
\begin{equation}
\hat{D}_{\eta _{\lambda }}^{\lambda }(\tau )=\chi (\tau \eta _{\lambda })
\hat{D}^{\lambda }(\tau ),  \label{Dnu}
\end{equation}
where different choices of $\eta _{\lambda }\in \mathbb{F}_{2^{n}}$
correspond to distinct relabelling of the eigenstates within each $\lambda $
-basis. A phase-space arrangement $\pi _{\mathbf{\eta }}$ (for a fixed
spread) is a choice of cocycle phases $\Phi _{\eta _{\lambda }}^{\lambda
}\left( \tau \right) \Leftrightarrow $ $\gamma _{\lambda }(\tau )$ at every
point of each curve in the partition:
\begin{equation}
\pi _{\mathbf{\eta }}=\{(\Gamma ^{\lambda },\Phi _{\eta _{\lambda
}}^{\lambda }),\;\lambda \in \Lambda \},\quad \mathbf{\eta }=(\eta _{\lambda
})_{\lambda \in \Lambda }.  \label{pinu}
\end{equation}

\begin{figure}[tbp]
\includegraphics[width=\linewidth]{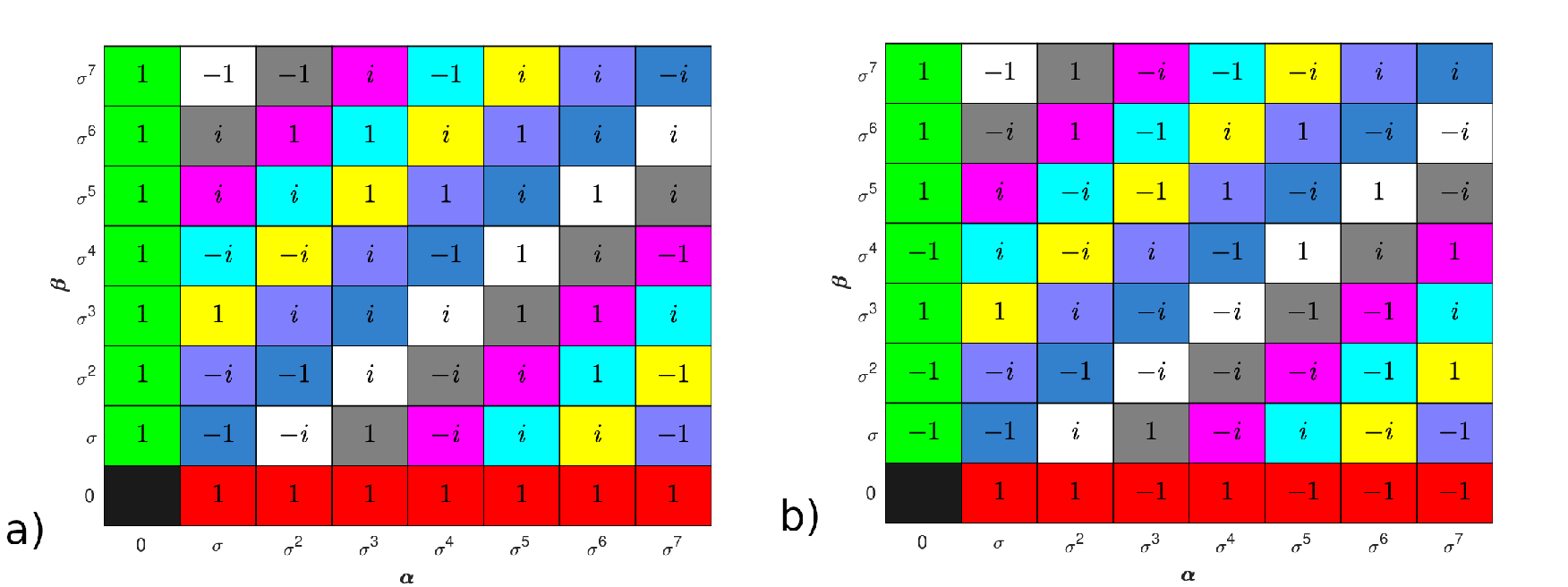}
\caption{a) Phase space arrangement corresponding to the partition (\ref{rays}) with all positive phases (\ref{Phil}); b) the same with
especially chosen signs of phases (\ref{Phinu}), $n=3$ \cite{data}.}
\label{fig:fig1ab}
\end{figure}

Since each of the $d+1$ curves admits $d$ gauge choices, a fixed partition
supports $d^{d+1}$ arrangements (quantum nets). In practice, it suffices to
set the signs of phases only at elements of the self-dual basis, $\Phi^{\lambda }\left( \tau =\theta _{i}\right) $, {$i=1,\ldots ,n,$ } the rest
of $\Phi ^{\lambda }\left( \tau \right) $ can be found by the recursion (\ref{Phi}) (see Appendix C). For instance, electing positive signs of the roots
of the structural equation (\ref{SE}),
\begin{equation}
\Phi ^{\lambda (+)}\left( \theta _{l}\right) =+\sqrt{\chi \left( \alpha
_{\lambda }(\theta _{l})\beta _{\lambda }(\theta _{l})\right) },
\label{Phil}
\end{equation}
we can parametrize all possible solutions of (\ref{Phi}) as in (\ref{Phinu}). Then, the sets $\{\eta _{\lambda }\}$, $\lambda \in \Lambda $ distinguish
the arrangements $\pi _{\eta }$, by associating eigenstates of the
stabilizers (\ref{Dg}) of unity eigenvalue with the parameters $\eta_{\lambda }$:
\begin{equation}
\hat{D}_{\eta }^{\lambda }(\tau )|\psi _{\eta _{\lambda }}^{\lambda }\rangle
=|\psi _{\eta _{\lambda }}^{\lambda }\rangle .  \label{Due}
\end{equation}
In Fig. \ref{fig:fig1ab} we show three-qubit DPS for the partition based
on rays (\ref{rays}) for two different arrangements, where a) corresponds to
all positive roots in (\ref{Phil}), and b) to a specifically chosen
arrangement, which will be later used for plotting Wigner functions of some
stabilizer states. An example of an arrangement for a partition locally
inequivalent to ({{\ref{rays}) }}is briefly discussed in Appendix C, Eqs. (\ref{090a})-(\ref{090d}).

In other words, fixing the cocycle gauge is the same as choosing which basis
state is assigned to the curve passing through the origin, and in general
ordering the elements of each basis  - curve (translate) correspondence:
\begin{equation}
v+\Gamma ^{\lambda }\Leftrightarrow |\psi _{\eta _{\lambda }+\sigma
_{\lambda }(v)}^{\lambda }\rangle \langle \psi _{\eta _{\lambda }+\sigma
_{\lambda }(v)}^{\lambda }|,\quad v\in V,  \label{psiG}
\end{equation}%
where $\sigma _{\lambda }(v)$ chooses a parallel curve, and $\{\eta
_{\lambda }\}$ fix the origin of the state labeling.

\section{Wigner functions of n-qubit stabilizer states}

\textit{Discrete Wigner representation and phase-space arrangements}

An operator $\hat{f}$ acting in $n$-qubit space $\mathcal{H}_{2^{n}}$ can be
represented as a real-valued distribution on the discrete phase space (DPS)
via a one-to-one Wigner map \cite{gibbons,gross,cormick,bjork2007,bjorkDWF,fasesqb,faseRauss,paz2005},
\begin{equation}
W_{f}\left( u\right) =Tr\left[ \hat{f}\hat{w}\left( u|\pi _{\mathbf{\eta }
}\right) \right] ,\quad u=\left( \alpha ,\beta \right) \in V,  \label{map}
\end{equation}
where $\hat{w}(u|\pi _{\mathbf{\eta }})$ is a Stratonovich-Weyl kernel
associated with a chosen phase-space partition $\pi $. A covariant under the
discrete displacements (\ref{Dg}) self-dual mapping kernel,
\begin{eqnarray}
\hat{w}\left( u|\,\pi _{\mathbf{\eta }}\right) &=&\hat{D}_{u}\hat{w}\left(
0|\,\pi _{\mathbf{\eta }}\right) \hat{D}_{u}^{\dagger },  \label{cov} \\
\hat{w}\left( 0|\pi _{\mathbf{\eta }}\right) &=&\sum_{\lambda \in \Lambda
}|\psi _{\eta _{\lambda }}^{\lambda }\rangle \langle \psi _{\eta _{\lambda
}}^{\lambda }|-\hat{I},  \label{w0c}
\end{eqnarray}
satisfying the standard requirements (see Appendix C), is entirely
determined by a phase-space arrangement $\pi _{\mathbf{\eta }}$ (\ref{pinu}). Using (\ref{label shift}) one can represent  the kernel (\ref{cov}) as a
sum of projectors onto the elements of the respective mutually unbiased
bases (\ref{UB}) \cite{gibbons,galvao,cormick,bjork2007,fasesqb,
bjorkDWF,paz2005}:
\begin{equation}
\hat{w}(u|\pi _{\mathbf{\eta }})=\sum_{\lambda }|\psi _{\eta _{\lambda
}+\sigma _{\lambda }(u)}^{\lambda }\rangle \langle \psi _{\eta _{\lambda
}+\sigma _{\lambda }(u)}^{\lambda }|-\hat{I}.  \label{wab}
\end{equation}%
Due to the freedom in choosing the phase-space arrangement, there are a
double-exponential number, $\left( 2^{n}\right) ^{2^{n}+1}$, of possible
maps (\ref{wab}).

This allows us to represent the Wigner function of a state $\hat{\rho}$
in terms of measured probabilities $p_{\lambda ,\kappa }=\langle \psi
_{\kappa }^{\lambda }|\hat{\rho}|\psi _{\kappa }^{\lambda }\rangle $ as
\begin{equation}
W_{\rho }(u|\pi _{\mathbf{\eta }})=\sum_{\lambda }p_{\lambda ,\eta _{\lambda
}+\sigma _{\lambda }(u)}-1,  \label{WFpsi}
\end{equation}%
and, conversely, leads to the tomographic expansion
\begin{equation}
\hat{\rho}=\sum_{\lambda ,\kappa }p_{\lambda ,\kappa }|\psi _{\kappa
}^{\lambda }\rangle \langle \psi _{\kappa }^{\lambda }|-\hat{I}.  \label{tom}
\end{equation}

Equivalently, the probabilities $p_{\lambda ,\kappa }$ are linked to the
Wigner marginals through
\begin{equation}
d^{-1}\sum_{u\in v+\Gamma ^{\lambda }}W_{\psi }(u|\pi _{\eta })=p_{\lambda
,\eta _{\lambda }+\sigma _{\lambda }(v)}.  \label{tomcond}
\end{equation}

Wigner kernels that satisfy the tomographic condition (\ref{tomcond}) for
different phase-space partitions are generally not equivalent under global
Clifford conjugation, even though the underlying sets of mutually unbiased
bases can be transformed into one another by Clifford operations \cite{Munoz17}.  This distinction arises because the discrete Wigner kernel
encodes not only the choice of MUBs but also their geometric and phase-space
assignment, and cocycle gauge, none of which are necessarily preserved
under Clifford equivalence alone.

\subsection{ State-adapted arrangements for stabilizer states}

Let us consider a stabilizer state $|s_{\Gamma ,\xi }\rangle $ determined by
a maximally isotropic curve $\Gamma $ in an arrangement (\ref{pinu}). Each $\lambda $-basis corresponds to a family
 of parallel phase-space curves $v+\Gamma ^{\lambda }$ according to (\ref{psiG}), where the curve $\Gamma
^{\lambda }$ (passing through the origin) is associated with the state $|\psi _{\eta _{\lambda }}^{\lambda }\rangle $. Among the $d$ curves parallel
to $\Gamma ^{\lambda }$, some intersect $\Gamma $ while others do not.
According to Proposition 1, the occupied labels of $|s_{\Gamma ,\xi }\rangle$
 in the $\lambda $-basis form the affine coset $\kappa _{\lambda}^{0}+H_{\lambda }$, where $H_{\lambda }=T_{\lambda }^{\perp }$, and $\kappa
_{\lambda }^{0}$ is fixed by the condition (\ref{chiz}). By choosing the
cocycle phases (\ref{Phinu}) so that $\eta _{\lambda }\in \kappa _{\lambda
}^{0}+H_{\lambda }$, Eq.~(\ref{crossG1}) yields the following equivalence
relations for any $v\in V$,

\begin{eqnarray}
(v+\Gamma ^{\lambda })\cap \Gamma &\neq &\varnothing \Leftrightarrow \sigma
_{\lambda }(v)\in H_{\lambda }  \label{GGl} \\
&\Leftrightarrow &\eta _{\lambda }+\sigma _{\lambda }(v)\in \kappa _{\lambda
}^{0}+H_{\lambda }.
\end{eqnarray}

Thus a $\lambda $-translate crosses $\Gamma $ exactly when the projector
assigned to it has nonzero overlap with the stabilizer state. Any such $\eta
_{\lambda }$ therefore selects a maximum-probability outcome,
\begin{equation}
p_{\lambda \eta _{\lambda }}^{(\Gamma )}=|\langle \psi _{\eta _{\lambda
}}^{\lambda }|s_{\Gamma ,\xi }\rangle |^{2}=|H_{\lambda }|^{-1}.
\label{plopt}
\end{equation}

Consequently, the rephased displacement operator (\ref{Dnu}) has eigenvalue $+1$ on $|s_{\Gamma ,\xi }\rangle $:
\begin{equation*}
\hat{D}_{\eta _{\lambda }}^{\lambda }(\tau )|s_{\Gamma ,\xi }\rangle
=|s_{\Gamma ,\xi }\rangle ,\quad \tau \in T_{\lambda }.
\end{equation*}

\subsection{Phase-space localization}

The preceding construction applies independently to all $d+1$ bases of the
partition.

\textit{Theorem 1 (optimized delta localization).}

Let $|s_{\Gamma ,\xi }\rangle $ be any $n$-qubit stabilizer state and let $\pi $ be any complete stabilizer-MUB partition of the DPS. In each MUB fix a
phase adjustment $\eta _{\lambda }^{opt}$ satisfying Eq. (\ref{chiz}), or,
equivalently, choose any maximum-probability outcome (\ref{plopt}) in that
basis:
\begin{equation}
\quad p_{\lambda \eta _{\lambda }^{opt}}^{(\Gamma )}=p_{\max }^{(\lambda
)}=|H_{\lambda }|^{-1}\text{ for all }\lambda ,  \label{OpC}
\end{equation}
thus associating MUB curves $\Gamma ^{\lambda }$ with the maximum-outcome
state $|\psi _{\eta _{\lambda }^{opt}}^{\lambda }\rangle $, and the Wigner
kernel (\ref{w0c}) takes the form
\begin{equation*}
\hat{w}\left( 0|\pi _{\mathbf{\eta }}\right) =\sum_{\lambda \in \Lambda
}|\psi _{\eta _{\lambda }^{opt}}^{\lambda }\rangle \langle \psi _{\eta
_{\lambda }^{opt}}^{\lambda }|-\hat{I}.
\end{equation*}

Then, in any such arrangement one has
\begin{equation}
W_{s_{\Gamma ,\xi }}^{\mathrm{opt}}(u|\pi )=W_{s_{\Gamma ,\xi }}^{\mathrm{opt
}}(u|\pi _{\mathbf{\eta }^{opt}})=\mathbf{1}_{\Gamma }(u),  \label{Dloc}
\end{equation}
that is, the optimized Wigner function equals one on the $d$ points of $\Gamma $ and zero elsewhere.

\textit{Proof.}

It follows from Eq. \ref{GGl} that the normalized sum of the geometric
indicator $\mathbf{1}_{\Gamma }$ along a translate $v+\Gamma ^{\lambda }$
for a fixed $\lambda $ is, according to (\ref{Tl}),
\begin{equation}
\frac{1}{d}\sum_{u\in v+\Gamma ^{\lambda }}\mathbf{1}_{\Gamma }(u)=\frac{
|\left( v+\Gamma ^{\lambda }\right) \cap \Gamma |}{d}=\frac{|T_{\lambda }|}{d
}\,\mathbf{1}_{T_{\lambda }^{\perp }}\!\left( \sigma _{\lambda }(v)\right) .
\label{1Gl}
\end{equation}

On the other hand, for a state-adapted assignment (\ref{psiG}) Eqs. (\ref{GGl})-(\ref{plopt}) give
\begin{equation}
p_{\lambda ,\eta _{\lambda }^{opt}+\sigma _{\lambda }(v)}=\frac{|T_{\lambda
}|}{d}\,\mathbf{1}_{T_{\lambda }^{\perp }}\!\left( \sigma _{\lambda
}(v)\right) .  \label{pSA}
\end{equation}

Therefore, the indicator $\mathbf{1}_{\Gamma }$ and the adapted Wigner
function have the same marginals on every line of every $\lambda $-striation:
\begin{equation}
\frac{1}{d}\sum_{u\in v+\Gamma ^{\lambda }}\mathbf{1}_{\Gamma
}(u)=p_{\lambda ,\eta _{\lambda }^{opt}+\sigma _{\lambda }(v)}=\frac{1}{d}
\sum_{u\in v+\Gamma ^{\lambda }}W_{s_{\Gamma ,\xi }}^{\mathrm{opt}}(u|\pi ).
\label{mm}
\end{equation}

Since the complete set of MUB probabilities uniquely determines the density
operator and a Wigner map is invertible, equality of all marginals
implies (\ref{Dloc}). By Proposition 1 we can carry out the optimization
procedure (\ref{OpC}) for any stabilizer state in an arbitrary partition.
The resulting optimized Wigner function is the same indicator $1_{\Gamma }$
for every partition.

In practice, in a fixed partition for a given state $|\psi \rangle $, we
compute (measure) the outcome probabilities $p_{\lambda \kappa }=|\langle
\psi _{\kappa }^{\lambda }|\psi \rangle |^{2}$ in every basis, keeping in
each step only the maximal probability ${p_{\lambda \eta _{\lambda }^{opt}}}$,
which fixes the seed $\lambda $-basis state ${|\psi _{\eta _{\lambda
}}^{\lambda }\rangle }$ needed for the construction of the mapping kernel (\ref{w0c}).
 Thus, no double-exponential search over arrangements is
required: the optimization is performed independently within each MUB of the
chosen partition. This simplification does not, however, imply polynomial
measurement complexity in $n$, since statistics from all $d+1$ MUBs are
still required.

The optimized phases are then computed as
\begin{equation}
\Phi _{\eta _{\lambda }}^{\lambda }\left( \tau \right) =\langle \psi _{\eta
_{\lambda }}^{\lambda }|\hat{Z}_{\alpha _{\lambda }(\tau )}\hat{X}_{\beta
_{\lambda }(\tau )}|\psi _{\eta _{\lambda }}^{\lambda }\rangle ^{\ast
},  \label{optPhi}
\end{equation}
for $\left( \alpha _{\lambda }(\tau ),\beta _{\lambda }(\tau )\right) \in
\Gamma ^{\lambda}$, determining the phase-space arrangement (\ref{pinu}).

Once the elements of MUBs, $\{|\psi _{\eta _{\lambda }}^{\lambda }\rangle $,
$\lambda \in \Lambda \}$, with nonzero overlaps with a given stabilizer
state are identified, the optimized kernel at zero is constructed as
specified in Eq. (\ref{w0c}). The Wigner kernel at any phase-space point
is then determined using the covariance property (\ref{cov}).

In odd prime-power dimensions, where the Wigner kernel is factorized by
particle, all pure stabilizer states are indicators of affine Lagrangian
subspaces in any covariant Wigner map. For qubits, the arrangement typically
must be adapted to the specific stabilizer state, except when it is already
a part of the arrangement, in which case the delta-form of the Wigner
function follows directly from Proposition 1. Thus, although no fixed qubit
discrete Wigner representation ensures nonnegativity of all stabilizer
states \cite{cormick,raussendorf17,cohomology}, Theorem 1 demonstrates that
a state-dependent cocycle gauge can recover the delta-function Wigner
representation for every stabilizer state. This result provides a structural
qubit analog of Hudson's theorem at the optimized level. In contrast to the
strong partition dependence of unoptimized Wigner functions, the optimized
distributions coincide across inequivalent phase-space partitions in any
state-adapted arrangement, thereby restoring the same $\delta $-localization
for a stabilizer state.

\begin{figure}[tbp]
\includegraphics[width=\linewidth]{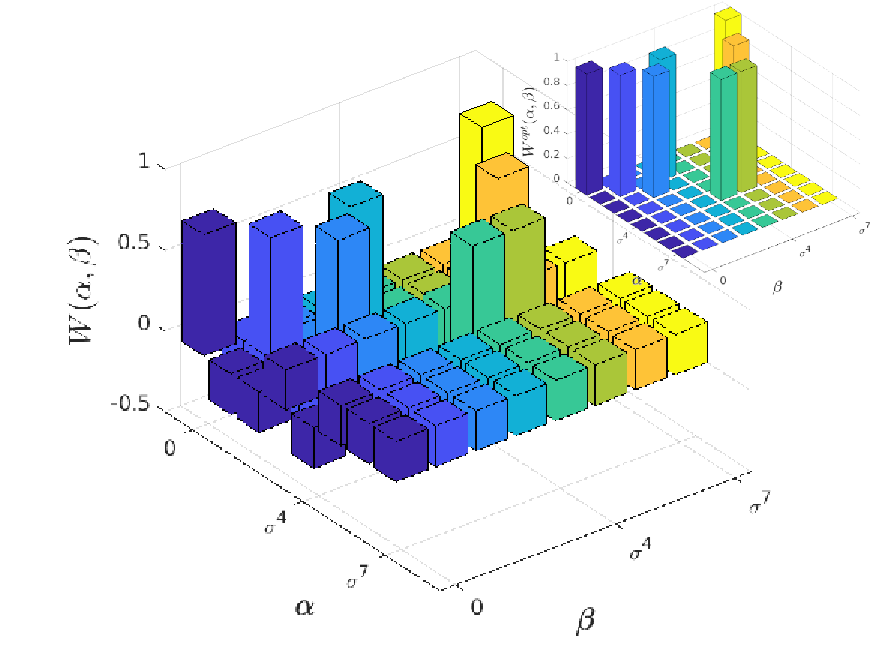}
\caption{a) The Wigner function for the {cluster } state (\ref{psimu}) in the arrangement given in Fig{. 1a }of the partition (\ref{rays}).
b) the same in the arrangement given in Fig. 1b, $n=3$ \cite{data}.}
\label{fig:fig2ab}
\end{figure}

In Fig. \ref{fig:fig2ab}, we plot the Wigner function of the closed
cluster state,
\begin{equation}
{\left\vert CS\right\rangle =2^{-n/2}\sum_{\tau
}(-1)^{\sum_{i=1}^{n}t_{i}t_{i+1}}|\tau \rangle ,\quad t_{n+1}=t_{1},\;}
\label{psimu}
\end{equation}
where ${\tau =}\sum_{i}\theta _{i}{t_{i}}$, ${t_{i}\in }\mathbb{Z}_{2}$,
corresponding {to }the commutative curve $\Gamma ^{c}=\{\left( \tau,\sum_{k}\left( \theta _{k+1}+\theta _{k-1}\right) tr\left( \theta _{k}\tau
\right) \right) $, $\tau \in \mathbb{F}_{2^{n}}\}$ \cite{curvas1}, in the
partition (\ref{rays}) and the arrangements presented in Fig. \ref{fig:fig1ab}. In the \textquotedblleft positive-phase\textquotedblright\
arrangement, a) the Wigner function is irregular, while the arrangement b)
selected by the optimization protocol (\ref{OpC}) produces the exact $\delta
$-shape (\ref{Dloc}) along $\Gamma ^{c}$. The optimized Wigner function for
the state (\ref{psimu}) in the partition (\ref{090a})-(\ref{090d}) has
exactly the same $\delta $-form (not shown).

\subsection{Flatness, phase matching, and the rigidity problem}

Let us apply the optimization protocol (\ref{OpC}) to an arbitrary pure
state $|\psi \rangle $. The optimized Wigner function at the origin is
\begin{equation}
W_{\psi }^{\mathrm{opt}}(0|\pi )=\sum_{\lambda \in \Lambda }p_{\max
}^{(\lambda )}-1,\quad p_{\max }^{(\lambda )}=\max_{\kappa }p_{\lambda
\kappa },  \label{Wmu1}
\end{equation}
where the maximization over outcome probabilities $p_{\lambda \kappa}=|\langle \psi _{\kappa }^{\lambda }|\psi \rangle |^{2}$ is carried out
independently in each $\lambda $-basis (with an arbitrary distribution of
phases $\Phi _{\eta _{\lambda }}^{\lambda }(\tau )$) used for the
construction of the partition. For a pure state, a complete set of $d+1$
MUBs is a complex projective two-design, satisfying the condition
\begin{equation}
\sum_{\lambda ,\kappa }p_{\lambda \kappa }^{2}=2,  \label{2d}
\end{equation}%
which along with the normalization $\sum_{\kappa }p_{\lambda \kappa }=1$
gives
\begin{equation}
W_{\psi }^{\mathrm{opt}}(0|\pi )-1=\sum_{\lambda ,\kappa }p_{\lambda \kappa
}\left( p_{\max }^{(\lambda )}-p_{\lambda \kappa }\right) \geq 0.  \label{W1}
\end{equation}

\textit{Lemma 1 (flatness bound).}

For every pure state and every complete stabilizer-MUB partition,
\begin{equation}
W_{\psi }^{\mathrm{opt}}(0|\pi )\geq 1,  \label{Wlb}
\end{equation}
where equality holds if and only if the probability distribution is flat on
its support in every MUB. In particular, every stabilizer state saturates
the bound.

Proof. Eq. (\ref{W1}) is a sum of nonnegative terms. Thus, equality in (\ref{W1})
 holds if and only if, in every basis, all nonzero probabilities equal
the corresponding maximum, $p_{\lambda \kappa }=p_{\max }^{(\lambda )}$. Let
us choose one occupied label $\kappa _{\lambda }^{0}$ in each basis and
shift the outcome labels so that it becomes zero. We denote the resulting
centered support by $H_{\lambda }$ and we use the same symbols for the
relabeled basis states. Thus, in any of mutually unbiased $\lambda $-basis a
state satisfying $W_{\psi }^{\mathrm{opt}}(0|\pi )=1$ has a flat
distribution,
\begin{equation}
|\psi \rangle =\frac{1}{\sqrt{|H_{\lambda }|}}\sum_{\kappa \in H_{\lambda
}}z_{\kappa }^{(\lambda )}|\psi _{\kappa }^{\lambda }\rangle ,\quad
|z_{\kappa }^{(\lambda )}|=1,  \label{psiF}
\end{equation}
with
\begin{equation}
p_{\lambda \kappa }=\frac{1}{|H_{\lambda }|}\mathbf{1}_{H_{\lambda }}(\kappa
),\quad \sum_{\lambda }\frac{1}{|H_{\lambda }|}=2,  \label{flats}
\end{equation}%
according to (\ref{2d}). It follows from Proposition~1 that every stabilizer
state is flat in all bases of the partition and therefore saturates the
bound $W_{\psi }^{\mathrm{opt}}(0|\pi )=1$ for any complete stabilizer MUB.

For any centered support, defining its annihilator by
\begin{equation*}
H_{\lambda }^{\perp }\equiv \{\tau :\func{tr}(h\tau )=0\text{ for all }h\in
H_{\lambda }\}=(\mathrm{span}_{\mathbb{F}_{2}}H_{\lambda })^{\perp },
\end{equation*}
Eq.~(\ref{SS1}) gives, for the original displacement operators,
\begin{equation}
\hat{D}_{\gamma _{\lambda }(\tau )}|\psi \rangle =\chi (\kappa _{\lambda
}^{0}\tau )|\psi \rangle ,\quad \tau \in H_{\lambda }^{\perp }.
\label{stabL}
\end{equation}
These operators share the eigenvector $|\psi \rangle $ and therefore
commute pairwise. Since distinct curves of the partition meet only at the
origin, their number is
\begin{equation}
N_{stab}=1+\sum_{\lambda \in \Lambda }(|H_{\lambda }^{\perp
}|-1)=\sum_{\lambda \in \Lambda }|H_{\lambda }^{\perp }|-d.  \label{Nstab}
\end{equation}
If $N_{stab}=d$, they form a maximal stabilizer group and $|\psi \rangle $
is a stabilizer state.

Equation~(\ref{stabL}) also implies that every centered $\lambda $-support
is invariant under the shifts generated by stabilizers obtained from any
other $\mu $-basis,
\begin{equation*}
H_{\lambda }=H_{\lambda }+\sigma _{\lambda }(\gamma _{\mu }(H_{\mu }^{\perp
})).
\end{equation*}
 The map $P_{\lambda \mu }=\sigma _{\lambda }\circ \gamma _{\mu }$ is an
additive isomorphism for $\lambda \neq \mu $, so $|P_{\lambda \mu }(H_{\mu
}^{\perp })|=|H_{\mu }^{\perp }|$.

\textit{Proposition 3 (low-dimensional rigidity and the first nonadditive
branch).}

(i) For one and two qubits, every all-MUB-flat pure state is a stabilizer
state. (ii) For arbitrary $n$, an all-MUB-flat pure state cannot have a
nonadditive centered support of cardinality three.

The proof is given in Appendix E.

The centered supports $H_{\lambda }$ and phases $z_{\kappa }^{(\lambda )}$
are constrained by the requirement that all expansions (\ref{psiF})
represent the same state. Let us choose one basis $\mu $ as a seed. For a
fixed centered seed support $H_{\mu }$, the continuous unknown variables in
a flat state $|\psi (\mathbf{z})\rangle $ are the relative phases $z_{\tau
}^{(\mu )}$, $|z_{\tau }^{(\mu )}|=1$, $z_{0}^{(\mu )}=1$. The period
subgroup of probabilities in any complementary $\lambda $-basis is $P_{\lambda \mu }(H_{\mu }^{\perp })\equiv B_{\lambda |\mu }$; hence the
probabilities in each complementary basis are constants on cosets of $B_{\lambda |\mu }$: $p_{\lambda \kappa }=p_{\lambda \kappa +b}$, $b\in
B_{\lambda |\mu }$, with $|B_{\lambda |\mu }|=2^{n-s}$, where $s=\dim _{\mathbb{F}_{2}}\mathrm{span}H_{\mu }$, $\mathbb{F}_{2^{n}}=\bigcup_{r\in
R_{\lambda }}\left( r+B_{\lambda |\mu }\right) $, and $R_{\lambda }$ is a
set of coset representatives. Therefore, the number of distinct outcome
probabilities in a $\lambda $-basis is $|\mathbb{F}_{2^{n}}/B_{\lambda |\mu
}|=2^{s}$, all of which are functions of $|H_{\mu }|-1$ seed phases $z_{\tau
}^{(\mu )}$, $\tau \in H_{\mu }\setminus \{0\}$. As a consequence, each
centered support $H_{\lambda }$ is a union of cosets
\begin{equation}
H_{\lambda }=\bigcup_{r\in J_{\lambda }}\left( r+B_{\lambda |\mu }\right) ,
\label{Hcoset}
\end{equation}%
where $r\in J_{\lambda }$ labels an occupied coset (one with nonzero
probability), $p_{\lambda r}(\mathbf{z})=|\langle \psi _{r}^{\lambda }|\psi (\mathbf{z})\rangle |^{2}\neq 0$, $r\in J_{\lambda }$, where $|J_{\lambda
}|=|H_{\lambda }|/|B_{\lambda |\mu }|$.

Choosing an occupied quotient label $r_{\lambda }^{0}$, $p_{\lambda
r_{\lambda }^{0}}(\mathbf{z})>0$, flatness in the $\lambda $-basis is
equivalent to a system of algebraic equations for $|H_{\mu }|-1$ seed phases
$\{z_{\tau }:\tau \in H_{\mu }\setminus \{0\}\}$:
\begin{equation}
p_{\lambda r}(\mathbf{z})\left( p_{\lambda r}(\mathbf{z})-p_{\lambda
r_{\lambda }^{0}}(\mathbf{z})\right) =0,\quad r\neq r_{\lambda }^{0}.
\label{Qflat}
\end{equation}

For the familiar Desarguesian MUB (\ref{MUBray})-(\ref{Vray}) the seed
support is convenient to choose as $H_{x}$, and the probabilities in the
complementary bases are
\begin{equation}
p_{\lambda \kappa }=\frac{1}{d|H_{x}|}|\mathcal{T}_{\lambda \kappa }(\mathbf{
z})|^{2},\quad \mathcal{T}_{\lambda \kappa }(\mathbf{z})=\sum_{\tau \in
H_{x}}\phi ^{\lambda }(\tau )^{\ast }\chi (\kappa \tau )z_{\tau }^{(\mu )},
\label{Tau}
\end{equation}
and these amplitudes are constant on cosets of $H_{x}^{\perp }$: $\mathcal{T}
_{\lambda ,\kappa +b}(\mathbf{z})=\mathcal{T}_{\lambda \kappa }(\mathbf{z})$
for $b\in H_{x}^{\perp }$.

Proposition~2 completely settles the additive branch: an all-MUB-flat state
with additive centered supports is a stabilizer state. Hence any
nonstabilizer flat solution, if it exists, must contain at least one
nonadditive centered support. Stabilizer states automatically satisfy the
system (\ref{Qflat}). The existence of nonadditive supports would require
exceptional phase cancellations that produce a non-subgroup selection of
equally weighted quotient cosets in every complementary MUB. Eqs. (\ref{Qflat}) test whether such nonadditive solutions can occur.

For each complementary basis, there are $2^{s}$ quotient labels. Thus,
selecting one pivot quotient outcome leaves $2^{s}-1$ equations in that
basis, and over the $d$ complementary bases the raw number of scalar
equations is
\begin{equation}
N_{\mathrm{eq}}(s)=d(2^{s}-1),  \label{Neq}
\end{equation}
where algebraic dependencies among these equations are not excluded. If $|H_{\mu }|=2^{s}$, the support is additive. Therefore, $s+1\leq |H_{\mu
}|\leq 2^{s}-1$, and $2^{s}-1$ is the largest possible nonadditive
cardinality for a fixed span dimension $s$. Consequently, for fixed $s$, the
smallest gap between the number of equations (\ref{Neq}) and the number of
independent phases (i.e., the overdetermination within the non-additive
branch) occurs for $|H_{\mu }|=2^{s}-1$:
\begin{equation}
\Delta (s)=d(2^{s}-1)-(2^{s}-2)=(d-1)(2^{s}-1)+1,  \label{gap}
\end{equation}%
and increases with $s$.

For $s=0$, the only centered support is $\{0\}$; for $s=1$, any centered
support spanning a one-dimensional binary space is the subgroup $\{0,a\}$.
The first possible nonadditive case occurs at $s=2$, where $H=\{0,a,b\}$,
with linearly independent $a,b \in \mathbb{F}_{2^{n}}$, and with two seed
phases satisfying $3d$ equations. Appendix E proves that this
cardinality-three nonadditive support is incompatible with all-MUB flatness.

The above exact results and extensive numerical simulations motivate the
following

\textit{Conjecture 1 (MUB flatness rigidity).}

For arbitrary $n$, a pure state that is flat on its support in every basis
of a complete stabilizer MUB is a stabilizer state. Equivalently, the set of
equations (\ref{Qflat}) does not admit nonadditive physical solutions, i.e.,
all supports of a flat state are additive subgroups (up to translation).

\begin{figure}[tbp]
\centering
\includegraphics[width=\linewidth]{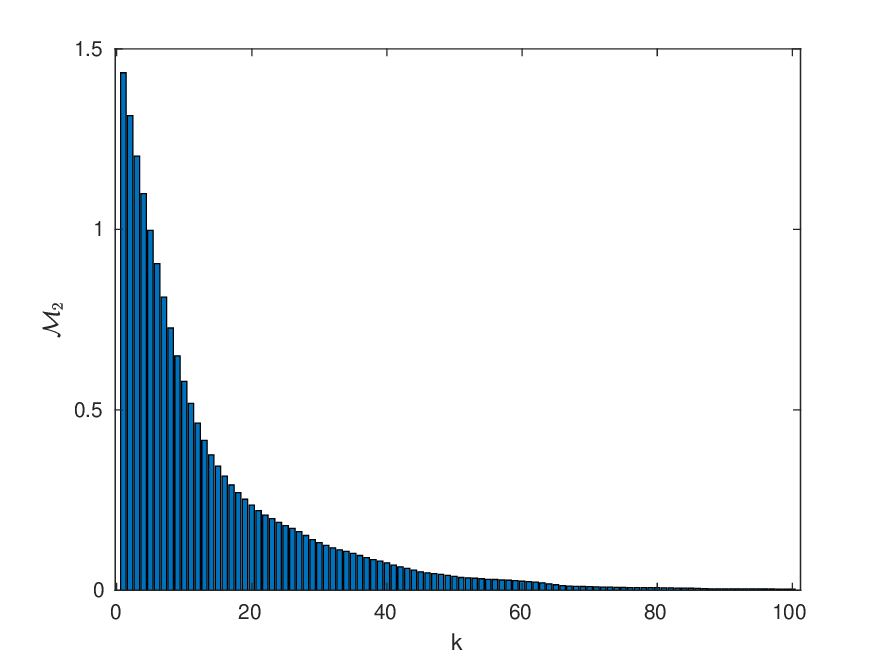}
\caption{The stabilizer entropy (\ref{R}) along numerical descent
toward flat states for five qubits. The bars are averages over 100
successful Haar-initialized trajectories that reached the minimum within
numerical precision in fewer than 120 accepted updates. The coordinate $k$
counts the number of steps used in the gradient optimization algorithm.}
\label{fig:Stab Ent}
\end{figure}

To test Conjecture~1 numerically, we monitored the Stabilizer R\'{e}nyi
Entropy (SRE) \cite{sre},
\begin{eqnarray}
\mathcal{M}_{2}&=&-\ln \left( d^{-1}\mathcal{R}(\hat{\rho})\right) ,\label{R}\\
\mathcal{R}(\hat{\rho})&=&\sum_{u}|r_{u}|^{4},\quad
 r_{u}=Tr\left( \hat{\rho}
\hat{D}_{u}\right) ,\nonumber
\end{eqnarray}
which vanishes for stabilizer states. We start with a random state and
employ the standard gradient method to decrease the value of the Wigner
function at the origin. In Fig. \ref{fig:Stab Ent} we plot $\mathcal{M}_{2}$
in each intermediate $k$-step averaged over 100 initial Haar states for $n=5$.
 The average entropy rapidly decays to zero as the state becomes flat. The
endpoint, where $\mathcal{M}_{2}=0$, always corresponds to a stabilizer
state. In this procedure, the supports of states, initially occupying the
whole field, shrink to an additive subspace that does not necessarily
coincide with $\mathbb{F}_{2^{n}}$.

\section{Non-stabilizer witness}

Localization Theorem 1 and Conjecture 1 suggest that the optimized Wigner
function at the origin (\ref{Wmu1}) characterizes the departure from the
flat-probability pattern. The optimization protocol {splits }into $d+1$
independent maximizations, one in each MUB, where only the largest
probability is stored in each step. The remaining phase-space values of the
Wigner function retain information about the chosen arrangement and will be
used mainly for visualization.

\subsection{One-sided witness and operational meaning}

The optimized value at the origin defines the partition-dependent score
\begin{equation}
M_{\pi }(\left\vert \psi \right\rangle )=W_{\psi }^{opt}(0|\pi ).  \label{M}
\end{equation}
The quantity $M_{\pi }(\left\vert \psi \right\rangle )$ admits a simple
operational interpretation: in each basis of the partition, one retains the
most likely outcome probability; summing these maximal probabilities over
all bases and subtracting one yields the optimized Wigner function at the
origin. Therefore, the existence of $d^{d+1}$ possible arrangements does not
imply a double-exponential search for the optimized origin value. Although
multiple optimization arrangements may exist, the value of $M_{\pi }$ is
unique. The optimization is restricted to arrangements within the selected
partition; evaluating $M_{\pi }$ does not require optimization over
different phase-space partitions. We stress that we do not need to
reconstruct the full Wigner function to determine the witness (\ref{M}).

According to Theorem 1, $M_{\pi }(\psi )>1$ implies that $|\psi \rangle $ is
non-stabilizer independent of $\pi $. In this sense, $M_{\pi }(\left\vert
\psi \right\rangle )$ quantifies how strongly $\left\vert \psi \right\rangle
$ deviates from the stabilizer pattern of probability concentration across a
complete set of mutually unbiased measurements. For $n\leq 2$ the converse
is proved above; for three qubits it is supported by explicit numerical
enumeration, while the first possible nonadditive support branch is excluded
in arbitrary dimension. Thus, $M_{\pi }$ is a one-sided witness detecting
MUB concentration: it rewards a state for having one large outcome
probability in each stabilizer basis. Conditional on Conjecture 1,
establishing the equivalence between $M_{\pi }(\psi )=1$ and stabilizerness,
$M_{\pi }(\psi )$ can be considered a genuine non-stabilizer witness.

\begin{figure}[tbp]
\includegraphics[width=\linewidth]{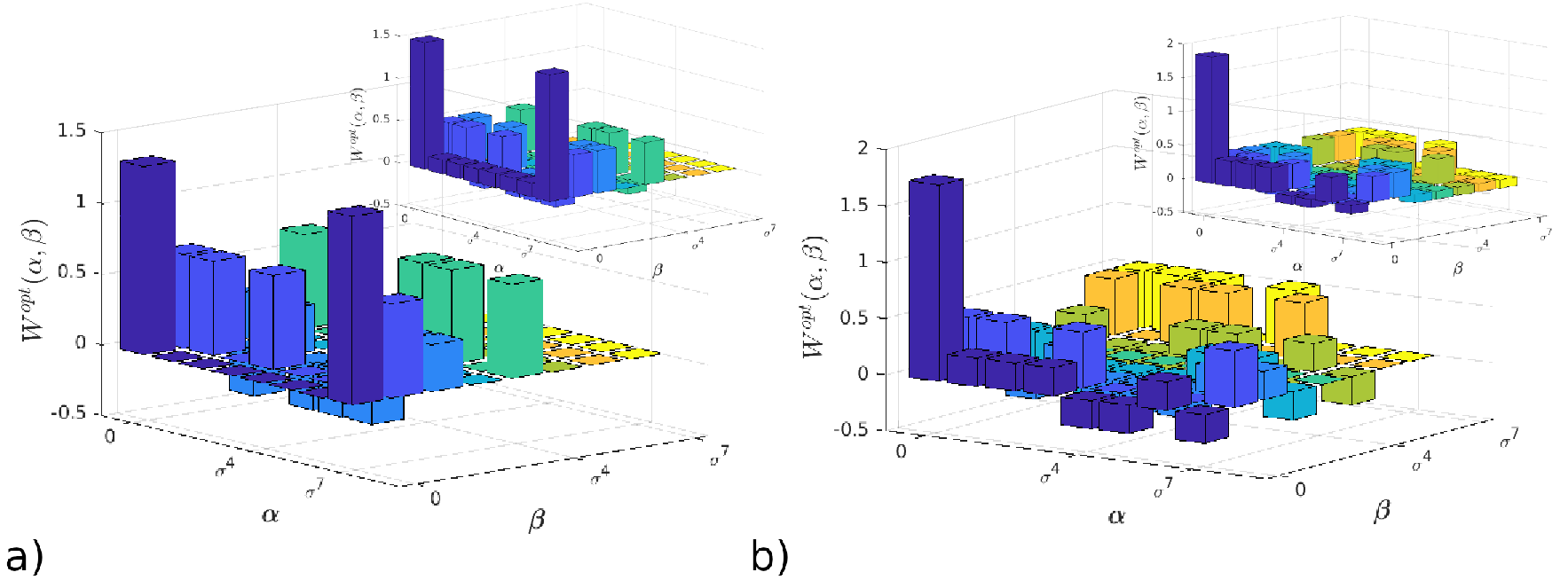}
\caption{The optimized Wigner function in the partition (\ref{rays})
for a) W-state, $|W\rangle \sim |100\rangle +|010\rangle +|001\rangle $; b)
domain wall state $|DW\rangle \sim |000\rangle +|100\rangle +|110\rangle
+|111\rangle $; Inset plots: the optimized Wigner functions for the same
states in the partition (\ref{090a})-(\ref{090d}), $n=3$
\cite{data}.}
\label{fig:fig3}
\end{figure}

In Fig. \ref{fig:fig3}, we plot the optimized Wigner functions of some
3-qubit non-stabilizer states in the partition (\ref{rays}). In the inset
plots, we show the same states in the partition (\ref{090a})-(\ref{090d})
for the corresponding optimized arrangements. One can observe some
similarity in the optimized Wigner functions across different partitions,
although in arbitrary arrangements, the Wigner functions have completely
different shapes.

For mixed states $\hat{\rho}$, the 2-design MUB identity
\begin{equation*}
\sum_{\lambda ,\kappa }p_{\lambda \kappa }^{2}=1+Tr\hat{\rho}^{2},\quad
p_{\lambda \kappa }=\langle \psi _{\kappa }^{\lambda }|\hat{\rho}|\psi
_{\kappa }^{\lambda }\rangle
\end{equation*}
along with the obvious inequality $p_{\max }^{(\lambda )}\geq \sum_{\kappa}p_{\lambda \kappa }^{2}$ gives
\begin{equation}
M_{\pi }(\hat{\rho})\geq \func{Tr}\hat{\rho}^{2}\geq d^{-1}.  \label{Mrho}
\end{equation}
However, since a maximum of linear functions is convex, $M_{\pi }(\hat{\rho})$
 is convex in $\hat{\rho}$. Thus, every state in the convex hull of pure
stabilizer states satisfies
\begin{equation}
M_{\pi }(\hat{\rho})\leq 1.  \label{Mrho1}
\end{equation}%
Consequently, $M_{\pi }(\hat{\rho})>1$ certifies that $\hat{\rho}$ lies
outside the stabilizer polytope.

A simple bound for the witness $M_{\pi }(\hat{\rho})$ follows immediately
from the expansion (\ref{lD}): since
\begin{equation*}
p_{\max }^{(\lambda )}=d^{-1}\left( 1+\max_{\kappa }\sum_{\tau \neq 0}\chi
(\kappa \tau )r_{\gamma _{\lambda }(\tau )}\right) ,
\end{equation*}
one has
\begin{equation}
M_{\pi }(\hat{\rho})\leq \mathcal{D}(\hat{\rho}),\quad \mathcal{D}(\hat{\rho}
)=d^{-1}\left( 1+\sum_{u\neq 0}|r_{u}|\right) ,  \label{MD}
\end{equation}
where $\boldsymbol{\mathcal{D}(\hat{\rho})}$ is the standard stabilizer
norm, bounded for pure states as
\begin{equation}
\mathcal{D}(\left\vert \psi \right\rangle )\leq r^{2}(\psi ),\quad |\psi
\rangle =\sum_{\tau }c_{\tau }|\tau \rangle ,  \label{Dbound}
\end{equation}
where $r(\psi )=\sum_{\tau }|c_{\tau }|$.

\subsection{Haar-random states: typical scale and concentration}

The witness (\ref{M}) is not monotone under free operations at finite system
size, but possesses several compelling asymptotic properties. We separate
two distinct problems: concentration for Haar-random states in a fixed
partition, and concentration of a fixed state sequence under random Clifford
transformations.

\medskip
\textit{Asymptotic behavior and Clifford invariance}

The asymptotic behavior of $M_{\pi }(\left\vert \psi \right\rangle )$ for
Haar-random states is governed by the statistics of the maximum component of
a Dirichlet$(1,...,1)$ distribution. Using the classical Darling-Holst
results \cite{Holst,Darling} one can prove (see Appendix F) the following

\textit{Theorem 2 (Haar concentration):} For Haar-random states $\left\vert \psi \right\rangle \in \mathcal{H}_{2^{n}}$, the non-stabilizer
witness (\ref{M}) in the asymptotic limit of a large number of qubits is
sharply concentrated around the partition-independent mean,
\begin{equation}
M_{Haar}=\mathbb{E}_{\mathrm{Haar}}M_{\pi }(\left\vert \psi
_{Haar}\right\rangle )=\frac{d+1}{d}H_{d}-1,  \label{MH}
\end{equation}
where $H_{d}=\sum_{m=1}^{d}m^{-1}$, so that asymptotically,
\begin{equation*}
M_{Haar}\approx n\ln 2+\gamma -1+O(n2^{-n}),\quad n\gg 1,
\end{equation*}
and
\begin{equation*}
\func{Var}_{\mathrm{Haar}}M_{\pi }(\left\vert \psi _{Haar}\right\rangle
)=O(n2^{-n/2}),  \label{VarH}
\end{equation*}
where $\gamma $\ is Euler's constant. Consequently, the witness (\ref{M})
is sharply concentrated for typical $n$-qubit states, $\lim_{n\rightarrow
\infty }\func{Var}_{\mathrm{Haar}}M_{\pi }(|\psi \rangle )=0$.

The Haar expectation in (\ref{MH}) is the same for every complete
MUB partition, since in each orthonormal basis the squared components
of a Haar-random state obey the same Dirichlet$(1,\ldots ,1)$ distribution.
Moreover, the variance
bound (\ref{VarH} is uniform in $\pi$, as its derivation relies solely on mutual-unbiasedness properties.
Therefore, for any two fixed complete MUB partitions
$\pi$ and $\pi'$, it holds
\begin{equation*}
M_{\pi}(|\psi_{\rm Haar}\rangle)
-
M_{\pi'}(|\psi_{\rm Haar}\rangle)
\xrightarrow{\rm Pr}0 ,
\end{equation*}
since both random variables concentrate around the same Haar mean.
\begin{figure}[tbp]
\includegraphics[width=\linewidth]{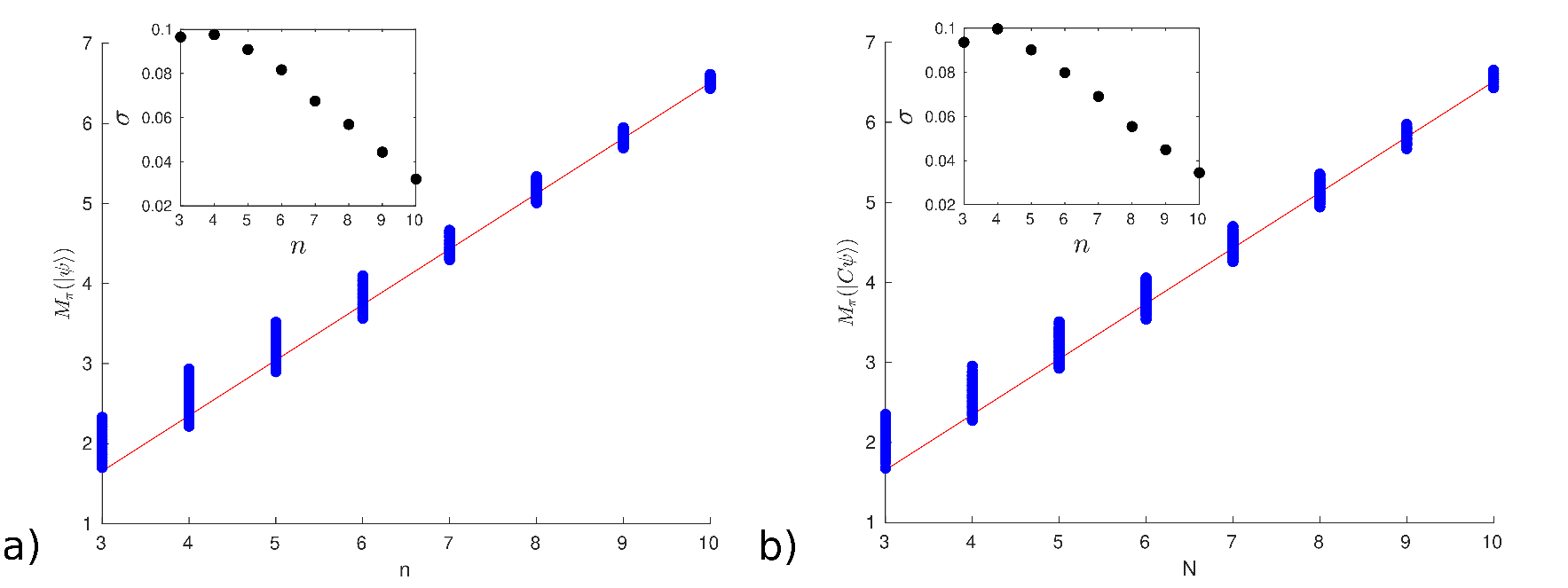}
\caption{a) The witness (\ref{M}) in the partition (\ref{rays}) as a function of the number of qubits for $1000$ Haar random
states. b) The witness (\ref{M}) in the partition (\ref{rays}) as a function of the number of qubits for a Haar state after $1000$
random Clifford transformations. Red line: $M=n\ln {2}+\gamma -1$
\cite{data}. }
\label{fig:fig4}
\end{figure}
In Fig. \ref{fig:fig4} we plot the witness value (\ref{M}): (a) for 1000
Haar-random states as a function of the number of qubits in the partition (\ref{rays});
(b) the same Haar state, transformed by $1000$ uniformly sampled Clifford unitaries, is shown. The progressive reduction of the spread
of the values of $M(\left\vert \psi _{Haar}\right\rangle )$ as $n$ grows
reflects the shrinking variance.

\textit{Corollary (Clifford orbit concentration).}

In the limit $n\gg 1$, the witness (\ref{M}) becomes asymptotically
invariant in distribution under random Clifford transformations
\begin{equation}
\lim_{n\rightarrow \infty }{\func{Var}}\left( M(\hat{C}\left\vert \psi
\right\rangle )\right) =0,\;\hat{C}\in \mathcal{C},  \label{VarMC}
\end{equation}
so that
\begin{equation*}
M(\hat{C}\left\vert \psi \right\rangle )\overset{\Pr }{\rightarrow }\mathbb{E
}_{C}M(\hat{C}\left\vert \psi \right\rangle ),
\end{equation*}
where the expectation and variance are taken over uniformly sampled
Clifford transformations. The witness concentrates around the
state-dependent Clifford-orbit mean, which need not coincide with the Haar
mean (\ref{MH}). The proof is given in Appendix G.

Eq. (\ref{VarMC}) shows that in the large-$n$ limit, different Clifford
realizations of the same state yield essentially the same witness values.
Equivalently, the witness of a given state becomes asymptotically
concentrated around the same mean for different partitions related by
Clifford transformations. This collapse of fluctuations justifies
interpreting (\ref{M}) as an asymptotic indicator of non-stabilizerness in
qubit systems.

\begin{figure}[tbp]
\includegraphics[width=\linewidth]{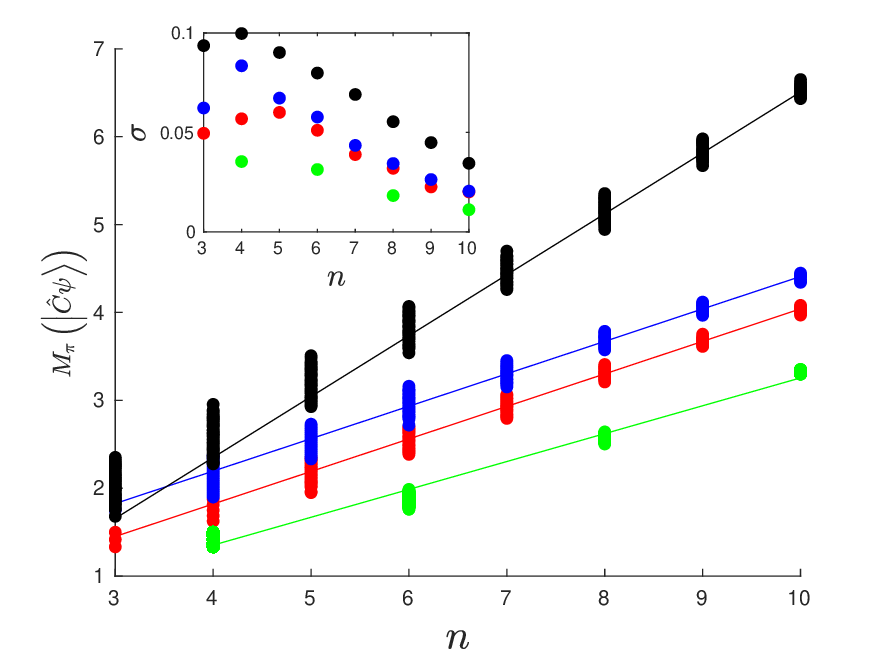}
\caption{The witness (\ref{M}) in the partition (\ref{rays})
for different states as a function of the number of qubits after $1000$
random Clifford transformations: Haar (black), sparse {$|W\rangle
\sim |100\ldots \rangle +|010\ldots \rangle +\cdots $} (red), and {$|DW\rangle \sim |000\ldots \rangle +|100\ldots \rangle +|110\ldots \rangle
+\cdots $} (blue) states, and the Dicke state $|k=n/2,n\rangle $ with
exponentially large computational support (green) \cite{data}. Inset
plot: the corresponding standard deviations $\sigma _{M}(n)$.}
\label{fig:fig5}
\end{figure}
In addition, for a wide class of states the witness behaves as
\begin{equation}
M_{\pi }(\hat{C}|\psi _{n}\rangle )=\Theta (n),  \label{Mn}
\end{equation}
In Appendix H we prove (\ref{Mn}) for sparse states whose computational
support $\mathbb{S}$ is affinely independent, and whose effective support $S_{\mathrm{eff}}(\psi )=(\sum_{\tau \in \mathbb{S}}|c_{\tau }|^{4})^{-1}$ is
$\Theta (n)$. This class includes, for example, the $W$ (one-excitation
Dicke) and domain-wall families. The numerical results indicate that linear
scaling persists for considerably broader $n$-qubit families $\{|\psi
_{n}\rangle \}$, including states with exponentially large computational
support. Appendix I gives a heuristic justification for this broader
behavior.

Eq. (\ref{Mn}), together with the numerical evidence for broader state
families, motivates the asymptotically linear scaling law

\begin{equation}
\mathbb{E}_{C}M(\hat{C}\left\vert \psi _{n}\right\rangle )\approx s_{\psi}n+o(n),  \label{MC}
\end{equation}
characterizing the rate of non-stabilizerness per qubit.

In Fig. \ref{fig:fig5}, we plot $M(\hat{C}\left\vert \psi \right\rangle )$
for several non-stabilizer states under 1000 random Clifford
transformations. All states exhibit an approximate linear dependence on the
number of qubits, in accordance with (\ref{MC}), with diminishing
fluctuations. This suggests the introduction of the asymptotic slope,
\begin{equation}
s_{\psi}=\lim_{n\rightarrow \infty }\frac{\mathbb{E}_{C}M(\hat{C}|\psi
_{n}\rangle )}{n},  \label{s}
\end{equation}
(whenever such limit exists) which provides a natural asymptotic
characterization of non-stabilizerness by capturing the rate at which it
grows with system size. For stabilizer states, $s=0$, whereas for
Haar-random states, $s=\ln 2$.

For several representative families the witness can be evaluated
analytically in the selected reference partition. In particular, $M(|W\rangle )=(n-1)/2+1/n$, and $M(|DW\rangle )=n/2+1/(n+1)$, so these
reference-partition values have asymptotic coefficient $1/2$, below the Haar
slope $\ln 2$. The numerical Clifford-orbit averages in Fig.~\ref{fig:fig5}
are consistent with $s_{\psi }=1/2$ for both families. For the SIC states
used in our numerical study \cite{scott}, the sampled values are fitted by $M(|SIC\rangle )\approx 0.559n+0.402$.

\begin{figure}[tbp]
\includegraphics[width=\linewidth]{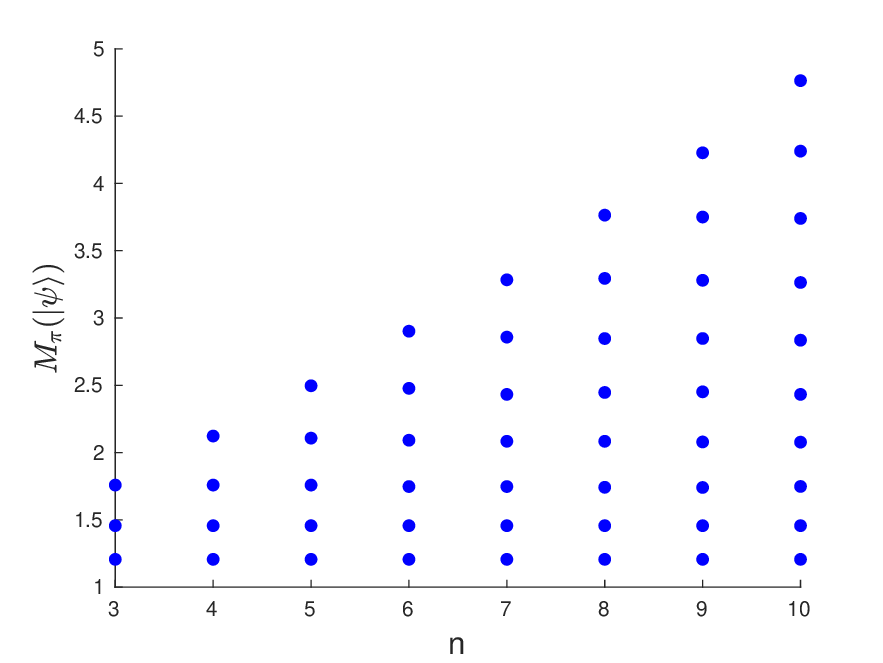}
\caption{The witness (\ref{M}) as a function of the number of qubits
for the cluster state (\ref{psimu}) after application of T-gates in
the partition (\ref{rays}). The vertical dots represent the number $
1\leq k\leq n$ of applied T-gates for each $n$ \cite{data}.}
\label{fig:fig6}
\end{figure}

\subsection{ Non-Clifford injection and scaling under T-gates}

It is instructive to examine how the measure (\ref{M}) changes when
non-Clifford gates are applied to a stabilizer state. In Fig. \ref{fig:fig6}
we plot $M_{\pi }(\left\vert \psi \right\rangle )$ for the cluster state (\ref{psimu}) in the arrangement (\ref{rays}) after a sequential application
of $1\leq k\leq n$ $T$-gates, $\hat{T}=diag(1,e^{i\pi /4})$, for different numbers of qubits.

For a fixed number of qubits, $M_{\pi }$ increases with the number of
applied $T$-gates, reflecting the progressive injection of non-Clifford
resources, while remaining strictly below the Haar benchmark $M_{\mathrm{Haar}}$; families with an extensive $T$-gate density display approximately
linear slopes according to (\ref{s}). We also observe a mild dependence of $M_{\pi }(\left\vert \psi \right\rangle )$ on the choice of the initial
stabilizer state: different stabilizer seeds lead to slightly different
slopes, but all exhibit the same  qualitative increase with non-Clifford
depth.

\subsection{ Maximally non-stabilizer states}

Within a fixed phase-space partition $\pi $, the maximally non-stabilizer
states according to the witness (\ref{M}) are eigenstates $|\varphi _{\max}\rangle $ of the Wigner kernels at the origin corresponding to the largest
eigenvalue $\varepsilon _{\max }$, i.e. $\varepsilon _{\max }\Leftrightarrow
|\varphi _{\max }\rangle $:
\begin{equation}
\hat{w}_{\max }\left( 0,0|\pi _{\eta _{\max }}\right) =\sum_{k}\varepsilon
_{k}|\varphi _{k}\rangle \langle \varphi _{k}|,  \label{wmax}
\end{equation}
so that, by construction,
\begin{equation*}
M_{\max }=\varepsilon _{\max }.
\end{equation*}
For a given partition, there may exist several arrangements $\pi _{\eta
_{\max }}$ leading to the same maximal eigenvalue $\varepsilon _{\max }$,
and there may exist several locally inequivalent $|\varphi _{\max }\rangle $.

Partitions related by a global Clifford transformation yield unitarily
equivalent Wigner kernels \textit{at zero}. Consequently, their maximizers
are Clifford-related and have the same $M_{\max }$.

Structurally, the mutually unbiased constraints favor $|\varphi _{\max}\rangle $ having a dominant component in each basis $\lambda $, while
suppressing the others. Since each MUB is a maximal commuting Pauli set,
fixing one dominant outcome per basis typically forces most nonidentity
Pauli expectations to remain relatively small.

As a concrete example, in the two-qubit case, $M_{\max }=1+\sqrt{3}/2$, and $|\varphi _{\max }^{(2)}\rangle =6^{-1/2}(\sqrt{3},i,1,-1)$ is one of
maximally non-stabilizer states in the partition (\ref{rays}). In the
three-qubit case, $M_{\max }=5/4+3\sqrt{3}/4$, and the state is $|\varphi
_{\max }^{(3)}\rangle \propto (-\xi ,i,1,\sqrt{3}\xi ,\xi ^{\ast },-\sqrt{3},\xi ,1)$, where $\xi =(\sqrt{3}-1)e^{-i\pi /4}/\sqrt{2}$. In addition, not
all kernels containing $\varepsilon _{\max }$ in their spectrum are
unitarily equivalent.

\begin{figure}[tbp]
\includegraphics[width=\linewidth]{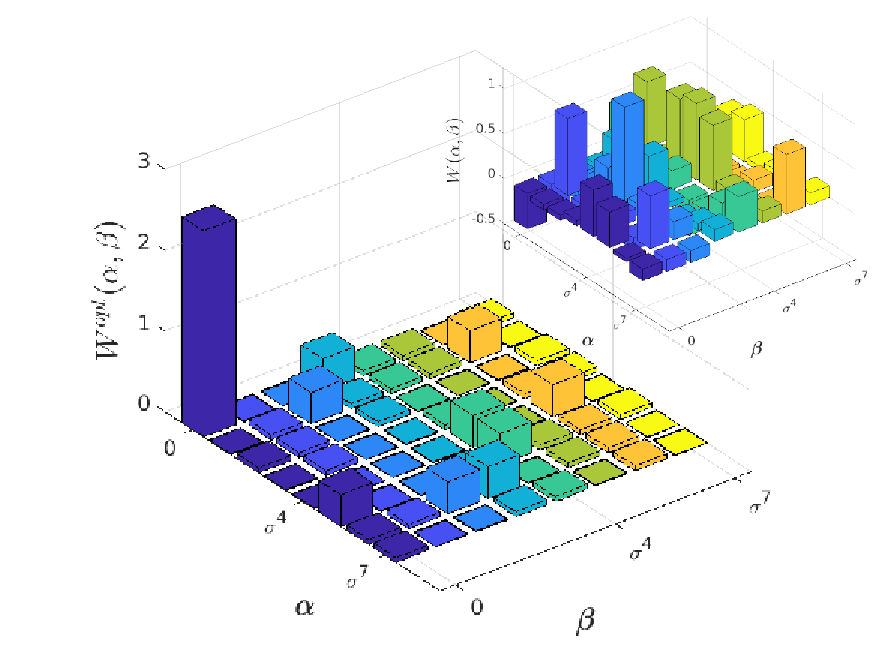}
\caption{The optimized Wigner function for the maximally non-stabilizer
state $|\varphi _{\max }\rangle $ in the partition (\ref{rays}); inset plot: the Wigner function of the state $|\varphi _{\max }\rangle $ in the arrangement given in Fig.~1a
\cite{data}.}
\label{fig:fig7}
\end{figure}

In Fig. \ref{fig:fig7}, we compare the optimized Wigner function and the
Wigner function in the arrangement given in Fig. \ref{fig:fig1ab} a) (inset
plot) of a maximally non-stabilizer three-qubit state (given above) in the
partition (\ref{rays}). The optimized Wigner function $W_{\varphi _{\max
}}^{opt}(\alpha ,\beta |\pi )$ exhibits a single sharp peak at the origin
and almost flat near-zero background in the rest of phase-space. In other
locally inequivalent partitions examined numerically, the optimized Wigner
functions display a similar qualitative structure and attain the same
maximal value at the origin.

It is worth noting that the optimized Wigner functions of some of $|\varphi
_{\max }\rangle $ in the two and three-qubit cases, in particular, above
mentioned $|\varphi _{\max }^{(2)}\rangle $ and $|\varphi _{\max
}^{(3)}\rangle $, are non-negative over the entire phase-space. These
states provide explicit examples with $Neg(\psi |\pi )=0$ while $M_{\pi
}(\psi )=M_{\max }>1$. This highlights that non-negativity alone could be
insufficient to characterize stabilizerness in qubit systems. In contrast,
exact delta-function localization emerges as the sharp geometric signature
distinguishing stabilizer states.

\subsection{ Comparison with the stabilizer R\'{e}nyi entropy}

\begin{figure}[tbp]
\includegraphics[width=\linewidth]{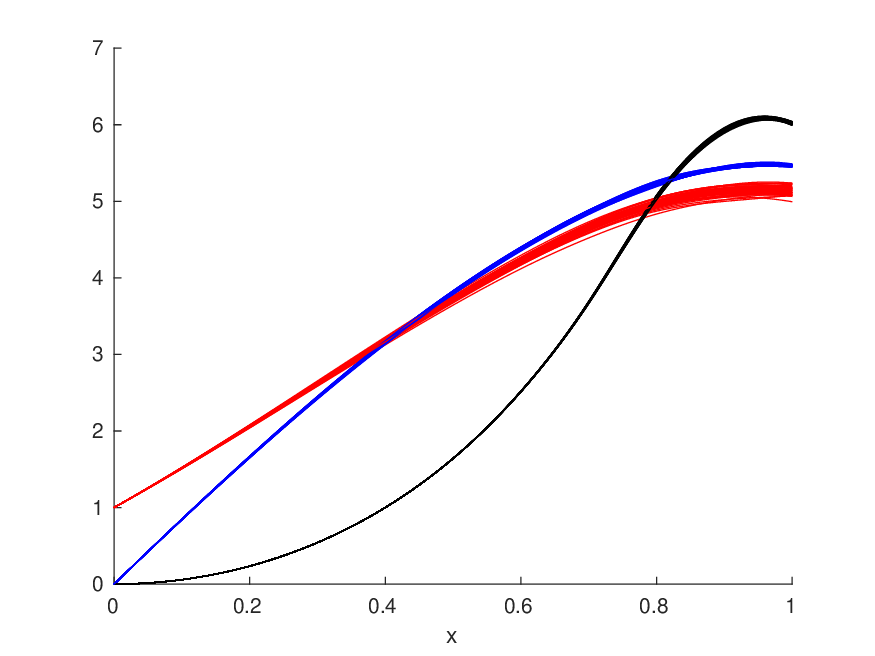}
\caption{Different measures of non-stabilizerness for $30$ states of the
form $\sqrt{1-x^{2}}|CS\rangle +x|H\rangle $, $\langle CS|H\rangle
=0 $: SRE (\ref{R}) (blue), the measure (\ref{M}) (red),
negativity (\ref{Neg}) (black), $n=8$ \cite{data}.}
\label{fig:fig9}
\end{figure}
The state-adapted optimization naturally suggests an alternative
diagnostic based on Wigner negativity. Motivated by the standard use of
Wigner negativity as a magic indicator \cite{veitch12}, we define the
state-adapted optimized negativity
\begin{equation}
Neg(\psi |\pi )=2^{-n-1}\sum_{u\in V}\left( |W^{opt}(u|\pi )|-W^{opt}(u|\pi
)\right) ,  \label{Neg}
\end{equation}
which vanishes for stabilizer states directly by Theorem~1. Numerical
results indicate that the fluctuations of $Neg(\psi )$ under sampled global
Clifford transformations decrease much less clearly than those of (\ref{M}).
We do not observe evidence for an analogous asymptotic Clifford-invariance
of (\ref{Neg}). Furthermore, the non-negativity of the optimized Wigner
function for some maximally non-stabilizer (and non-flat) states $|\varphi
_{\max }\rangle $ for two and three qubits shows that the negativity $Neg(\psi )$ may fail to detect non-stabilizerness \cite{faseRauss}.

On the other hand, the Stabilizer R\'{e}nyi Entropy (\ref{R}), which is
Clifford invariant by construction, in the Wigner representation has the form
\begin{equation*}
\mathcal{M}_{2}=-\ln \left[ d^{-3}\sum_{\mu ,\nu }\left( \sum_{\alpha ,\beta
}W(\alpha ,\beta )W(\alpha +\mu ,\beta +\nu )\right) ^{2}\right] ,
\end{equation*}
and detects global phase-space correlations rather than localized structure
as $M(\left\vert \psi \right\rangle )$.

As shown in Fig. \ref{fig:fig9}, $\mathcal{M}_{2}$, $M(\left\vert \psi
\right\rangle )$, and $Neg(\psi )$ display qualitatively similar behavior
along interpolations between stabilizer and Haar-random states, yet select
different extremal states. In particular, SIC states maximizing SRE \cite{magicSIC}, are maximally non-stabilizer in the sense of isotropic
delocalization: their stabilizer measurement statistics are as uniform as
mutual unbiasedness allows. In contrast, the witness (\ref{M}) rewards
anisotropic concentration, favoring states for which each mutually unbiased
basis contains a single dominant outcome. These two notions of
non-stabilizerness are therefore conceptually distinct.

\begin{figure}[tbp]
\includegraphics[width=\linewidth]{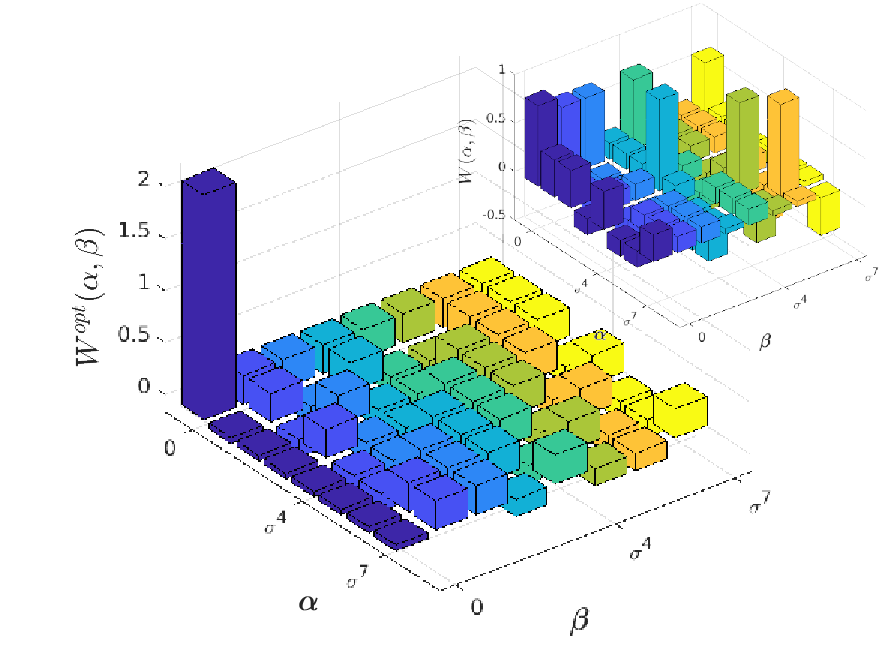}
\caption{The optimized Wigner function for the SIC-state 8a, taken from \cite{scott} in the partition (\ref{rays}) Inset plot: the Wigner function in the arrangement given
in Fig.1a., $n=3$ \cite{data}.}
\label{fig:fig10}
\end{figure}

Despite this distinction, the optimized Wigner functions of SIC states and
of the maximizers $|\varphi _{\max }\rangle $ exhibit a remarkably similar
geometry: both display a single pronounced peak at the origin with
low-amplitude fluctuations elsewhere in phase-space, see Fig. \ref{fig:fig10}
. Quantitatively, however, SIC states do not saturate the maximal eigenvalue
of the optimized Wigner kernel. We consistently observe $M(\left\vert
SIC\right\rangle )<M_{\max }$, where $M_{\max }$ is the spectral maximum of
\ the optimized Wigner kernel: for two, three, and four qubits, $M(\left\vert SIC\right\rangle )$ is respectively $1.5709$,
 $2.1666$, and $2.4984$; in comparison,
 for states $|\varphi _{\max }\rangle $ one has $M_{\max }=1.8660$, $2.5490$, and $3.4257$. Thus, whether (\ref{R}) and (\ref{M}) capture complementary or partially overlapping resources remains an
open question.

\section{Conclusions}

In odd-dimensional systems, discrete Wigner functions offer a clear
geometric picture of stabilizer states and quantum magic. In contrast,
discrete phase-space methods for qubit systems are considered intrinsically
limited by the incompatibility among covariance, Wigner-function
factorization, and stabilizer positivity. This limitation has hindered the
development of a transparent physical interpretation of stabilizer
classicality in experimentally relevant settings. The apparent pathologies
of qubit discrete Wigner functions arise not from inherent flaws in
phase-space methods, but from the restriction to fixed, state-independent
arrangements. Allowing the phase-space construction to adapt to the state
through an optimization procedure reveals a stabilizer-specific geometric
structure that otherwise remains hidden.

For any fixed phase-space partition, every qubit stabilizer state can be
represented as a discrete Wigner distribution in the form of a
delta-function supported on the corresponding stabilizer curve, provided the
phase-space arrangement is adjusted to the state. This localization property
is not intrinsic to the partition alone, but emerges only after an
appropriate selection of cocycle phases. Notably, although the underlying
gauge freedom of qubit discrete phase space ensures the existence of such
arrangements, the optimization can be performed directly using
experimentally accessible measurement statistics by selecting the outcome
with maximal probability in each mutually unbiased basis (MUB) of the
partition. This avoids an exhaustive search over all possible phase-space
arrangements.

Our probability-based construction provides a practical route to
state-adapted phase-space representations and establishes a qubit analogue
of Hudson-type extremal characterization: although no rigidly constructed
discrete Wigner function can characterize stabilizer states by positivity
alone, every stabilizer state admits a delta-function representation once
the phase-space arrangement is optimized. It should be stressed that, unlike
Hudson's theorem in odd dimensions \cite{gross}, our construction is
intrinsically state-dependent and should be viewed as its structural
counterpart, rather than a universal phase-space description, and therefore
does not contradict the existing no-go theorems \cite{raussendorf17,catani,catani18,cohomology} for qubit discrete Wigner
functions. This objective differs from that of framed Wigner functions \cite{park}, which exploit additional frame freedom primarily to extend classical
simulability.

Within this framework, we introduce a witness of non-stabilizerness (\ref{M}),
defined as the optimized value of the Wigner function at the origin,
operationally corresponding to the sum of maximal probabilities across all
mutually unbiased bases of a partition, minus unity. Stabilizer states
saturate the minimal value of this witness $M_{\pi }=1$, while $M_{\pi }>1$
ensures non-stabilizerness for pure states; for mixed states, $M_{\pi
}(\rho )>1$ certifies that $\hat{\rho}$ lies outside the stabilizer
polytope. The deviation from the minimum of $M_{\pi }$ quantifies the
anisotropic concentration across a complete set of MUBs. For Haar-random
states, the witness is sharply concentrated around the mean $n\ln
2+\gamma -1+o(1)$, which provides the natural typical-state benchmark for
large $n$. Although $M_{\pi }=1$ guarantees only all-MUB-flatness, specific
examples, numerical simulations, and the increasing overdetermination count (\ref{gap}) support the conjecture that only stabilizer states attain this
minimum.

Since the optimization procedure is state-dependent, the witness $M_{\pi}(\left\vert \psi \right\rangle )$
is not a resource monotone at finite
system size. However, as the number of qubits increases, the witness (\ref{M})
becomes asymptotically invariant in probability under random Clifford transformations, suggesting the introduction of a linear growth law (\ref{MC})
and an associated non-stabilizer rate per qubit (\ref{s}). We support this
emergent invariance analytically through the Clifford-orbit concentration
bound and numerically through random Clifford sampling for a broad class of
non-stabilizer families.

Overall, our results show that the $n$-qubit discrete phase-space can
support a meaningful geometric characterization of stabilizer structure and
provide an operational diagnostic of non-stabilizerness based on
experimentally accessible statistics. The present approach connects
algebraic and geometric perspectives on qubit magic, complementing existing
stabilizer-entropy-based measures, and reveals qualitatively new anisotropic
features of non-stabilizerness.

\acknowledgments

This work is partially supported by the Grant CBF2023-2024-50 of SECIHTI
(Mexico).

\appendix

\section{Stabilizer states in the $X$ basis}

The general form of a stabilizer state in the basis $\{|\tilde{\tau}\rangle\}$
 of eigenstates of the operators $\hat{X}_{\nu }$, $\hat{X}_{\nu }|\tilde{\tau}\rangle =\chi (\nu \tau )|\tilde{\tau}\rangle $ is
\begin{equation}
|\psi _{\kappa }^{\lambda }\rangle =2^{-q\boldsymbol{_{\lambda }}
/2}\sum_{\tau \in H_{\lambda ,\kappa }}\chi (h_{\lambda }(\kappa )\tau )\phi
^{\lambda }\left( \tau \right) |\tilde{\tau}\rangle ,  \label{psiX}
\end{equation}
where $H_{\lambda ,\kappa }$ is an affine coset of an additive subgroup
$H_{\lambda }\subset \mathbb{F}_{2^{n}}$, all such supports having
cardinality $|H_{\lambda }|=2^{q_{\lambda }}$, $0\leq q_{\lambda }\leq n$.
The phases $\phi ^{\lambda }(\tau )$  satisfy the cocycle condition
obtained by applying Eq. (\ref{DZX}) to Eq. (\ref{psiX}) and comparing
expansion coefficients,
\begin{equation}
\phi ^{\lambda }\left( \alpha _{\lambda }(\tau )+\tau ^{\prime }\right)
=\phi ^{\lambda }(\alpha _{\lambda }(\tau ))\phi ^{\lambda }(\tau ^{\prime
})\chi \left( \beta _{\lambda }(\tau )\tau ^{\prime }\right) ,
\label{coscond}
\end{equation}
and related to $\Phi ^{\lambda }\left( \tau \right) $ in (\ref{DZX}) through $\Phi ^{\lambda }\left( \tau \right) =\phi ^{\lambda }\left( \alpha
_{\lambda }(\tau )\right) $; the function $h_{\lambda }(\kappa )$, $h_{\lambda }(0)=0$, is a linear labeling map, chosen consistently with
the completeness of the basis in Eq.~(\ref{psiX}), and is defined from
the condition $tr(h_{\lambda }(\kappa )\alpha _{\lambda }(\tau ))=tr(\kappa
\tau )$, $\tau \in H_{\lambda }$ with $\{\alpha _{\lambda }(\tau ),\beta
_{\lambda }(\tau )\}$ denoting the elements of the $\Gamma ^{\lambda }$-set.

An analogous expansion holds in the computational basis, with the roles of $\alpha _{\lambda }(\tau )$ and $\beta _{\lambda }(\tau )$ exchanged.

\bigskip

\section{Proof of Proposition 1}

In this Appendix we prove Proposition 1 (stabilizer flatness).

It follows from the normalization condition
\begin{equation*}
\sum_{\kappa \in \mathbb{F}_{2^{n}}}|\langle \psi _{\kappa }^{\lambda
}|s_{\Gamma ,\xi }\rangle |^{2}=1,
\end{equation*}%
that {at } least one overlap $\langle \psi _{\kappa }^{\lambda }|s_{\Gamma
,\xi }\rangle $  in (\ref{sk}) is nonzero. For such a $|\psi _{\kappa
}^{\lambda }\rangle $, the matrix element of a displacement operator
labelled by $\tau \in T_{\lambda }$, can be computed in two ways:
\begin{align}
{\langle \psi _{\kappa }^{\lambda }|\hat{D}_{\gamma _{\lambda }(\tau
)}|s_{\Gamma ,\xi }\rangle }& {=\chi _{\xi }(\gamma _{\lambda }(\tau
))\langle \psi _{\kappa }^{\lambda }|s_{\Gamma ,\xi }\rangle ,} \\
& {=\chi (\kappa \tau )\langle \psi _{\kappa }^{\lambda }|s_{\Gamma ,\xi
}\rangle .}
\end{align}
Nonvanishing overlap implies the compatibility equations
\begin{equation}
\chi (\kappa \tau )=\chi _{\xi }(\gamma _{\lambda }(\tau )),\quad \tau \in
T_{\lambda }.  \label{chiz1}
\end{equation}
If $\kappa _{\lambda }^{0}$ is a solution of (\ref{chiz1}), then every
occupied label belongs to the affine coset $\kappa _{\lambda
}^{0}+H_{\lambda }$, where $H_{\lambda }=T_{\lambda }^{\perp }$.

To determine which labels are occupied we compute the trace of the product
of the projector expansions
\begin{align}
|s_{\Gamma ,\xi }\rangle \langle s_{\Gamma ,\xi }|& =d^{-1}\sum_{u\in \Gamma
}\chi _{\xi }(u)\hat{D}_{u},  \label{sD} \\
|\psi _{\kappa }^{\lambda }\rangle \langle \psi _{\kappa }^{\lambda }|&
=d^{-1}\sum_{\tau \in \mathbb{F}_{2^{n}}}\chi (\kappa \tau )\hat{D}_{\gamma
_{\lambda }(\tau )},  \label{lD}
\end{align}
which is non-zero only for $\gamma _{\lambda }(\tau )\in \Gamma \cap \Gamma
^{\lambda }$. The character sum appearing in
\begin{equation*}
p_{\lambda \kappa }^{(\Gamma )}=d^{-1}\sum_{\tau \in T_{\lambda }}\chi
(\kappa \tau )\chi _{\xi }(\gamma _{\lambda }(\tau )),
\end{equation*}
equals $|T_{\lambda }|$ for every label in $\kappa _{\lambda
}^{0}+H_{\lambda }$ and zero otherwise. Therefore, all labels in this coset
are occupied with probability $|T_{\lambda }|/d$, proving Eq.~\ref{sk}.

Thus, $H_{\lambda }=T_{\lambda }^{\perp }$ is the centered support. In the
original basis labeling,
\begin{equation*}
|s_{\Gamma ,\xi }\rangle =|H_{\lambda }|^{-1/2}\sum_{h\in H_{\lambda
}}z_{h}|\psi _{\kappa _{\lambda }^{0}+h}^{\lambda }\rangle ,\quad |z_{h}|=1,
\end{equation*}%
where $|H_{\lambda }|=d/|T_{\lambda }|$, so that the state $|s_{\Gamma ,\xi
}\rangle $ is an eigenstate of $\hat{D}_{\gamma _{\lambda }(\tau )}$, for
all $\tau \in T_{\lambda }$. In particular, the dimension of a stabilizer
support in the computational basis $|H_{stab}|=2^{n-r_{z}}$, where $r_{z}$
is the number of diagonal stabilizers in the corresponding stabilizer set,
since each $\hat{Z}_{\alpha }$ operator imposes one linear constraint on
the expansion coefficients.

From the geometrical perspective, if $u\in \Gamma $ and $\tau \in T_{\lambda}$,
 then $\gamma _{\lambda }(\tau )\in \Gamma $, and since $\Gamma $ is
isotropic,
\begin{equation*}
0=[u,\gamma _{\lambda }(\tau )]=\func{tr}(\sigma _{\lambda }(u)\tau ),
\end{equation*}
so that, $\sigma _{\lambda }(u)\in T_{\lambda }^{\perp }=H_{\lambda }$, and
thus, $\sigma _{\lambda }(\Gamma )\subseteq H_{\lambda }$. The kernel of $\sigma _{\lambda }$ restricted to $\Gamma $,
\begin{equation}
\ker (\sigma _{\lambda }|_{\Gamma })=\Gamma \cap \ker \sigma _{\lambda
}=\Gamma \cap \Gamma ^{\lambda },  \label{kerG}
\end{equation}
is of cardinality
\begin{equation}
|\sigma _{\lambda }(\Gamma )|=\frac{|\Gamma |}{|\Gamma \cap \Gamma ^{\lambda
}|}=\frac{d}{|T_{\lambda }|}=|T_{\lambda }^{\perp }|=|H_{\lambda }|.
\label{Psize}
\end{equation}
The inclusion and the same size of both sides in (\ref{Psize}) imply
\begin{equation}
\sigma _{\lambda }(\Gamma )=T_{\lambda }^{\perp }=H_{\lambda }.
\label{sigmaT1}
\end{equation}

\section{Three-qubit inequivalent partition and phase-space arrangement}

As an example, let us consider intersections between curves of two different
partitions in the three-qubit case: rays (\ref{rays}) and a complete set of
MUB curves locally inequivalent to (\ref{rays}) and parametrically defined
by
\begin{eqnarray}
\alpha _{\mu }\left( \tau \right) &=&\tau \left( \sigma +\mu \right) +\tau
^{2}\left( \sigma ^{5}+\sigma ^{2}\mu ^{2}\right)  \notag \\
&&+\tau ^{4}\left( \sigma ^{6}+\sigma \,\mu ^{4}\right) ,  \label{090a} \\
\beta _{\mu }\left( \tau \right) &=&\tau \left( \sigma ^{3}+\sigma \mu
\right) +\tau ^{2}\left( \sigma ^{5}\mu ^{2}+\sigma ^{3}\right)  \notag \\
&&+\tau ^{4}\left( \sigma ^{6}\mu ^{4}+\sigma ^{5}\right) ,\quad \mu \in
\mathbb{F}_{2^{3}},  \label{090b}
\end{eqnarray}
and
\begin{eqnarray}
\alpha \left( \tau \right) &=&\tau ^{4}\sigma +\tau ^{2}\sigma ^{2}+\tau ,
\label{090c} \\
\beta \left( \tau \right) &=&\tau ^{4}\sigma ^{6}+\tau ^{2}\sigma ^{5}+\tau
\sigma ,  \label{090d}
\end{eqnarray}
where $\sigma $ is a root of the equation $1+\sigma +\sigma ^{3}=0$. The
curves (\ref{090a})-(\ref{090d}) are mutually disjoint, and fill the DPS;
the corresponding eigenstates determine a complete set of MUBs locally
inequivalent to (\ref{rays}), characterized by a bi-partite factorization of
all bases \cite{curvas,lawrence,romero07}.

The intersection condition with rays $(\alpha _{\lambda }\left( \tau \right)
=\tau ,\beta _{\lambda }\left( \tau \right) =\lambda \tau )$ for a fixed $\lambda $ and $\mu $ is just
\begin{equation*}
\beta _{\mu }\left( \tau \right) =\lambda \alpha _{\mu }\left( \tau \right) .
\end{equation*}

\begin{figure}[tbp]
\includegraphics[width=\linewidth]{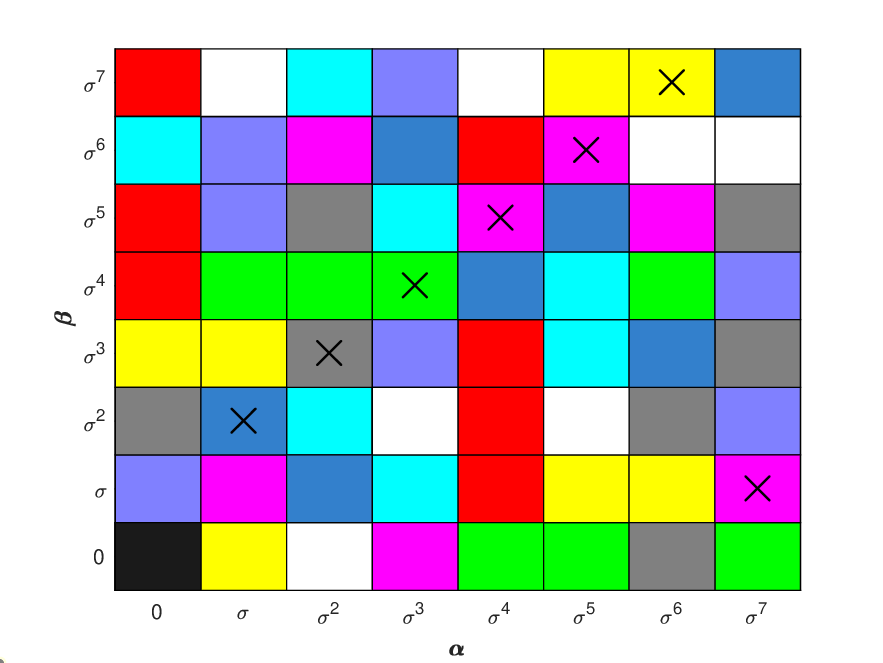}
\caption{Phase-space partition corresponding to the curves (\ref{090a}-\ref{090d}).
The crossed line is the ray $\beta =\sigma\alpha $, which intersects once each of the curves (\ref{090a}-\ref{090b})
 with $\mu =\sigma^{2},\sigma ^{5},\sigma ^{7}$, three times the curve with $\mu =\sigma^{4}$, and once the curve (\ref{090c},\ref{090d}), $n=3$ \cite{data}.}
\label{fig:fig11}
\end{figure}

In Fig. \ref{fig:fig11}, we show the phase-space arrangement optimized for
the cluster state (\ref{psimu}) and corresponding to the partition (\ref{090a})-(\ref{090d}).

For completeness, we give a closed-form solution of the cocycle recursion (\ref{Phi}) in a self-dual basis; it expresses the total phase as the product
of basis phases $\Phi _{\alpha ,\beta }^{\lambda }\left( \theta _{l}\right) $
times the accumulated commutation character,
\begin{eqnarray}
&&\Phi _{\alpha ,\beta }^{\lambda }\left( \tau
=\sum\limits_{k=1}^{n}t_{k}\theta _{k}\right) =  \label{Phi G} \\
&&\chi \left( \sum\limits_{m=1}^{n-1}\alpha _{\lambda }(t_{m}\theta
_{m})\beta _{\lambda }\left( \sum\limits_{k=m+1}^{n}t_{k}\theta _{k}\right)
\right) \prod\limits_{l=1}^{n}\left( \Phi _{\alpha ,\beta }^{\lambda }\left(
\theta _{l}\right) \right) ^{t_{l}},  \notag
\end{eqnarray}

being $\{\theta _{l}$, $l=1,...,n\}$ elements of the self-dual basis in $\mathbb{F}_{2^{n}}$, for each curve $\Gamma ^{\lambda }=\{(\alpha _{\lambda
}\left( \tau \right) ,\beta _{\lambda }\left( \tau \right) )\}$ in a given
partition.

A covariant under the discrete displacements (\ref{Dg}) Hermitian Wigner
mapping kernel $\hat{w}\left( u\right) =\hat{w}^{\dagger }(u)$, $u=\left(
\alpha ,\beta \right) $
\begin{equation}
\hat{w}\left( u\right) =\hat{D}_{u}\hat{w}\left( 0\right) \hat{D}
_{u}^{\dagger },  \label{cov1}
\end{equation}
is normalized according to,
\begin{equation}
\sum_{u}\hat{w}\left( u\right) =2^{n}\hat{I},\quad Tr\hat{w}\left( u\right)
=1,
\end{equation}
and satisfies the overlap relation,
\begin{equation}
Tr\left( \hat{w}(u)\hat{w}(u^{\prime })\right) =2^{n}\delta _{u\,u^{\prime
}}.
\end{equation}

Wigner kernels in different arrangements of the same partition are related
through \cite{faseqb1},
\begin{eqnarray}
\hat{w}(\alpha ,\beta |\pi _{\eta }) &=&2^{-n}\sum_{\nu ,\nu ^{\prime }}\hat{
w}(\alpha +\nu ,\beta +\eta \left( \nu ^{\prime }\right) +\nu ^{\prime }\nu
|\pi _{\eta =0})  \notag \\
&&+2^{-n}\sum_{\nu }\hat{w}(\alpha ,\nu |\pi _{\eta =0})-\hat{I},
\label{wnu}
\end{eqnarray}
where $\pi _{\eta =0}$ indicates the arrangement with all positive signs of
the phases (\ref{Phil}).

\section{Projection of a stabilizer curve into the Desarguesian MUB}

In this Appendix we discuss in detail how an arbitrary stabilizer curve $\Gamma $ is seen from rays of the Desarguesian MUB.

For instance, for the Desarguesian MUB (\ref{DM1})-(\ref{DM2}), where $\gamma _{x}(\tau )=(0,\tau )$, $\gamma _{\lambda }(\tau )=(\tau ,\lambda
\tau )$, taking $u=(\alpha ,\beta )$ we have for the vertical ray,
\begin{equation*}
\func{tr}(\sigma _{x}(u)\tau )=[u,\gamma _{x}(\tau )]=\func{tr}(\alpha \tau
)\Rightarrow \sigma _{x}(\alpha ,\beta )=\alpha ,
\end{equation*}
while for a finite ray,
\begin{eqnarray*}
\func{tr}(\sigma _{\lambda }(u)\tau )&=&[u,\gamma _{\lambda }(\tau )]\\
&=&\func{tr}((\beta +\lambda \alpha )\tau )\Rightarrow \sigma _{\lambda }(\alpha ,\beta
)=\beta +\lambda \alpha ;
\end{eqnarray*}
thus, for the projections of {points }of a ray $\mu \neq \lambda $, we
obtain
\begin{equation*}
\sigma _{\lambda }(\gamma _{\mu }(\tau ))=\sigma _{\lambda }(\tau ,\mu \tau
)=(\lambda +\mu )\tau ,
\end{equation*}
which is an isomorphism.

For a stabilizer curve $\Gamma $ which is not a ray, we find by (\ref{sigmaT}
) the following supports,
\begin{equation}
H_{x}=\{\alpha :(\alpha ,\beta )\in \Gamma \},\quad H_{\lambda }=\{\beta
+\lambda \alpha :(\alpha ,\beta )\in \Gamma \},  \label{derRaySupports}
\end{equation}
and due to (\ref{Tl})
\begin{equation}
H_{x}^{\perp }=\{\tau :(0,\tau )\in \Gamma \},\quad H_{\lambda }^{\perp
}=\{\tau :(\tau ,\lambda \tau )\in \Gamma \}.  \label{derRayAnnihilators}
\end{equation}

On the other hand, if $\Gamma =\Gamma ^{\lambda _{0}}$, then one has $
|T_{\lambda _{0}}|=d$ and $|H_{\lambda _{0}}|=1$, and if $\lambda \neq
\lambda _{0}$, so that $\Gamma \cap \Gamma ^{\lambda }=\{0\}$, then $
|T_{\lambda }|=1$ and $|H_{\lambda }|=d$, i.e., a ray stabilizer has one
delta-supported MUB distribution and is completely flat in all complementary
ray bases.

Note that $\ker \left( \sigma _{x}|_{\Gamma }\right) =\{(0,\beta ),\beta \in
H_{x}^{\perp }\}$, meaning that if $(\tau ,\beta )\in \Gamma $, then $(\tau
,\beta )+(0,\beta ^{\prime })=(\tau ,\beta +\beta ^{\prime })\in \Gamma $
for any $\beta ^{\prime }\in H_{x}^{\perp }$. Thus, the set of stabilizing
phase-space points with horizontal projection $\alpha =\tau $ is given by
the fiber {$\left( \sigma _{x}|_{\Gamma }\right) ^{-1}(\tau )=\{(\tau ,\beta
)\in \Gamma \}$}, $\tau \in H_{x}$. Any other point in the same fiber
differs by $(0,\beta ^{\prime })$, which lives in the kernel, and therefore $
|\sigma _{x}^{-1}(\tau )|=|\ker \left( \sigma _{x}|_{\Gamma }\right)
|=|H_{x}^{\perp }|$, i.e., all points of $H_{x}$ have the same horizontal
degeneracy. Similarly, $|H_{\lambda =0}^{\perp }|$ determines the vertical
degeneracy, i.e., the multiplicity of stabilizing phase-space points with
the same projection onto the $\beta $ axis.

\section{Proof of Proposition 3}

In this Appendix we prove Proposition 3.

A centered support has trivial annihilator if and only if it spans the whole
$n$-dimensional binary label space,
\begin{equation*}
H_{\lambda }^{\perp }=\{0\}\quad \Longleftrightarrow \quad \mathrm{span}_{
\mathbb{F}_{2}}H_{\lambda }=\mathbb{F}_{2^{n}}.
\end{equation*}
In this case $H_{\lambda }$ must contain at least $n$ linearly independent
nonzero labels, so that $|H_{\lambda }|\geq n+1$. If every annihilator is
trivial, then
\begin{equation}
2=\sum_{\lambda =1}^{2^{n}+1}\frac{1}{|H_{\lambda }|}\leq \frac{2^{n}+1}{n+1}.  \label{bound1}
\end{equation}
Since for $n=1$, and $n=2$ the right-hand sides are $3/2$ and $5/3$
respectively, at least one support has a nontrivial annihilator, and thus,
a non-identity stabilizer.

For $n=1$, any nontrivial Pauli stabilizer generates a maximal isotropic
subgroup of size $2$, which implies that the state is a stabilizer state.

For $n=2$, Eq. (\ref{bound1}) implies that at least one nonzero stabilizer
exists. Suppose that the stabilizer subspace is one-dimensional, $
\{0,u\}\subset \Gamma ^{\lambda _{0}}$, and in that basis $|H_{\lambda
_{0}}^{\perp }|=2$, and therefore, $\dim _{\mathbb{F}_{2}}\mathrm{span}
H_{\lambda _{0}}=1$, so that $|H_{\lambda _{0}}|=2$. For $\mu \neq \lambda
_{0}$, the same stabilizer $u\notin \Gamma ^{\mu }$ produces a nonzero label
translation $a_{\mu }=\sigma _{\mu }(u)\neq 0$, such that $H_{\mu }+a_{\mu
}=H_{\mu }.$ Thus, $H_{\mu }$ is a union of cosets of $\{0,a_{\mu }\}$, and
for two qubits a nonempty centered invariant support can therefore have
size $2$ or $4$. If $|H_{\mu }|=2$, then $H_{\mu }=\{0,a_{\mu }\}$, which
is additive subgroup of dimension one. In this case, $H_{\mu }^{\perp }$ has
order two and contains a nonzero element, yielding another nonidentity
stabilizer on $\Gamma ^{\mu }$, independent of $u$. This contradicts the
assumed uniqueness of the nonidentity stabilizer. Therefore, the only
remaining possibility is $|H_{\mu }|=4$, $\mu \neq \lambda _{0}.$The support
sizes are then $2,4,4,4,4$, and the reciprocal sum in (\ref{flats}) is $3/2$.
 This excludes a one-dimensional stabilizer subspace in the two-qubit case.
Since a nontrivial stabilizer has already been shown to exist, the
stabilizer subspace must therefore have dimension two; it is maximal
isotropic, and $|\psi \rangle $ is a stabilizer state.

Consider now an $n$-qubit all-MUB-flat state with a nonadditive centered
support in the $\mu $-th basis of the form $H=\{0,a,b\}$, where $a,b \in
\mathbb{F}_{2^{n}}$ are linearly independent. Selecting this basis as the
seed $\mu $-basis $\{|\psi _{\kappa }^{\mu }\rangle \}$ yields:
\begin{equation*}
|\psi (\mathbf{z})\rangle =\frac{1}{\sqrt{3}}\left( |\psi _{0}^{\mu }\rangle
+z_{a}^{(\mu )}|\psi _{a}^{\mu }\rangle +z_{b}^{(\mu )}|\psi _{b}^{\mu
}\rangle \right) .
\end{equation*}
In this case, $\dim _{\mathbb{F}_{2}}\func{span}\{a,b\}=2$, and therefore $
|H_{\mu }^{\perp }|=d/4$; the corresponding seed-basis displacements keep
the state invariant. In every complementary basis $\lambda \neq \mu $, the
period subgroup $B_{\lambda |\mu }$ has size $d/4$. The probabilities in
each $\lambda $-basis are constant on the cosets of $B_{\lambda |\mu }$,
with only four different quotient outcomes, whose unnormalized amplitudes (\ref{Tau}) are of the form
\begin{equation}
\mathcal{T}_{\lambda \kappa }(\mathbf{z};\epsilon ,\eta )=1+\epsilon
\mathcal{A}_{\lambda }(\mathbf{z})+\eta \mathcal{B}_{\lambda }(\mathbf{z}
),\quad \epsilon ,\eta \in \{+1,-1\},  \label{4A}
\end{equation}
where $|\mathcal{A}_{\lambda }|=|\mathcal{B}_{\lambda }|=1$, due to mutual
unbiasedness. These factors $\mathcal{A}_{\lambda },\mathcal{B}_{\lambda }$
contain the fixed MUB transition phases multiplied by the same two seed
phases $z_{a},z_{b}$, and therefore, are not independent from one
complementary basis to another. For instance, for a Desarguesian partition
(\ref{MUBray}) taking $x$-basis as a seed, one has
\begin{equation*}
\mathcal{T}_{\lambda \kappa }(\mathbf{z};\epsilon ,\eta )=1+\phi ^{\lambda
}\left( a\right) ^{\ast }\chi (a\kappa )z_{a}^{(\mu )}+\phi ^{\lambda
}\left( b\right) ^{\ast }\chi (b\kappa )z_{b}^{(\mu )}.
\end{equation*}%
The corresponding intensities are
\begin{equation}
|\mathcal{T}_{\lambda \kappa }(\mathbf{z};\epsilon ,\eta )|^{2}=3+2\epsilon
x+2\eta y+2\epsilon \eta z,  \label{T2}
\end{equation}
where $x=\func{Re}\mathcal{A}_{\lambda }$, $y=\func{Re}\mathcal{B}_{\lambda }$,
$z=\func{Re}(\mathcal{A}_{\lambda }\mathcal{B}_{\lambda }^{\ast })$.

At most one amplitude $\mathcal{T}_{\lambda \kappa }(\mathbf{z};\epsilon,\eta )$ can vanish.
 If two vanishing amplitudes differ in only one sign,
subtracting their equations would give $\mathcal{A}_{\lambda }=0$ or $\mathcal{B}_{\lambda }=0$, which contradicts the requirement of unit
modulus. If the amplitudes differ in both signs, adding the equations
results in $2=0$. Therefore, at least three quotient amplitudes must be
nonzero. Furthermore, the four nonzero intensities (\ref{T2}) cannot be
equal, as this equality would impose the contradictory conditions $x=y=z=0$
(if $\mathcal{A}_{\lambda },\mathcal{B}_{\lambda }\in \{i,-i\}$ then $z=\pm 1$).

Flatness requires all nonzero quotient intensities to be equal. Given at
most one amplitude can vanish and four equal nonzero intensities are
impossible, each complementary basis must contain exactly one absent and
three occupied quotient classes. Each of three quotient classes has size $d/4 $. Therefore, according to (\ref{Hcoset}), $|H_{\lambda \neq \mu }|=3d/4$, and the relation (\ref{flats}) gives
\begin{equation*}
\sum_{\lambda \in \Lambda }\frac{1}{|H_{\lambda }|}=\frac{1}{3}+d\,\frac{1}{3d/4}=\frac{5}{3}\neq 2.
\end{equation*}
This contradiction shows that a flat state cannot have the nonadditive
centered support $H=\{0,a,b\}$ in any of its MUBs. Since any basis can be
chosen as the seed, no support of an all-MUB-flat state can realize this
nonadditive pattern.

An analogous exclusion analysis can be carried out for three qubits, but we
do not include that derivation here. Independently, our numerical
enumeration of three-qubit flat-support solutions found only stabilizer
states.

\section{Proof of Haar concentration}

In this Appendix, we prove Theorem 2, which states that for Haar-random
states the witness (\ref{M}) is sharply concentrated in the limit $n\gg 1$
with $M_{\pi }(\left\vert \psi _{Haar}\right\rangle )\rightarrow n\ln
2+\gamma -1+O(n2^{-n})$, and {{$\func{Var}_{\mathrm{Haar}}M_{\pi
}(\left\vert \psi _{Haar}\right\rangle )\rightarrow 0$}}.

\textit{Proof.} Let $\left\vert \psi \right\rangle \in \mathcal{H}_{d}$ be a
Haar-random state, $d=2^{n}$. In each basis $\{|\psi _{\kappa }^{\lambda
}\rangle $, $\kappa \in \mathbb{F}_{2^{n}}\}$ of a given partition $\pi $
the probabilities $p_{\lambda \kappa }=|\langle \psi _{\kappa }^{\lambda
}|\psi \rangle |^{2}$ obey joint Dirichlet($1,...,1$) distribution, in
particular, each marginal satisfies
\begin{equation}
p_{\lambda \kappa }\sim \text{Beta}(1,d-1),\quad \Pr (p_{\lambda \kappa
}>t)=(1-t)^{d-1}.  \label{pkt}
\end{equation}
The expectation value of the maximum probability $p_{\max }^{(\lambda
)}=\max_{\kappa }p_{\lambda \kappa }$ in each basis in the limit of large
dimensions satisfies the conditions \cite{Holst} (maximum stick length and
its variation),
\begin{eqnarray}
\mathbb{E}(p_{\max }^{(\lambda )}) &=&\frac{H_{d}}{d}\approx \frac{\ln
d+\gamma }{d}+O(d^{-2}),  \notag \\
{\func{Var}}(p_{\max }^{(\lambda )}) &=&\frac{\pi ^{2}/6+o(1)}{d^{2}},
\label{Emax}
\end{eqnarray}
here $H_{d}=\sum_{m=1}^{d}m^{-1}$, and $\gamma $ is Euler's constant.
Moreover, $p_{\max }^{(\lambda )}$ {in }a single basis is distributed
according to the Gumbel law \cite{Darling,Holst} in the limit $d\gg 1$
\begin{eqnarray}
\Pr (p_{\lambda \kappa } &>&t)\sim d^{-1}e^{-x},\;t=\frac{\ln d+x}{d},
\label{pkA} \\
\Pr (p_{\max }^{(\lambda )} &>&t)\sim 1-\exp \left( -e^{-x}\right) .
\label{pmax}
\end{eqnarray}
Thus, the distribution of $p_{\max }^{(\lambda )}$ is concentrated within a
narrow band of width $\delta t\sim d^{-1}$ around $t=d^{-1}\ln d$, the
so-called Darling window. Since a partition $\pi $ contains $d+1$ mutually
unbiased bases, linearity yields,
\begin{eqnarray*}
\mathbb{E}\left[ M_{\pi }(\left\vert \psi _{Haar}\right\rangle )\right]
&=&\sum_{\lambda }\mathbb{E}\left[ p_{\max }^{(\lambda )}\right] -1=(d+1)
\frac{H_{d}}{d}-1 \\
&=&\ln d+\gamma -1+O(\ln d/d),\quad d\gg 1,
\end{eqnarray*}
which is the mean-value statement of Theorem 2.

In order to prove the concentration of $M_{\pi }(\left\vert \psi
_{Haar}\right\rangle )$ we use the spherical-Poincar\'{e} inequality stating
that on a high-dimensional sphere, a function cannot fluctuate strongly if
its average squared rate of change along the sphere is small.

The function
\begin{equation}
F_{\pi }(\psi )=M_{\pi }(\psi )+1=\sum_{\lambda }\max_{\kappa }\langle \psi |
\hat{\Pi}_{\lambda \kappa }|\psi \rangle ,  \label{Fpi}
\end{equation}
where $\hat{\Pi}_{\lambda \kappa }=|\psi _{\kappa }^{\lambda }\rangle
\langle \psi _{\kappa }^{\lambda }|$, can be considered as a finite sum of
maxima of smooth quadratic functions on the real unit sphere $S^{2d-1}\subset \mathbb{C}^{d}\simeq $ $\mathbb{R}^{2d}$ with real inner
product $g(u,v)=\func{Re}\langle u|v\rangle $. However, $F_{\pi }(\psi )$ is
not smooth at points where two or more probabilities in the same basis
attain the maximum. Nevertheless, since the probabilities satisfy
\begin{eqnarray}
|p_{\lambda \kappa }(\psi )-p_{\lambda \kappa }(\phi )| &\leq &|\langle \psi
-\phi |\Pi _{\lambda \kappa }|\psi \rangle |+|\langle \phi |\Pi _{\lambda
\kappa }|\psi -\phi \rangle |  \notag \\
&\leq &2\Vert \psi -\phi \Vert ,
\end{eqnarray}
and $p_{\lambda \kappa }(\phi )\leq p_{\max }^{(\lambda )}(\phi )$,
choosing a maximizing $\kappa $ for $\psi $ and then exchanging $\psi $ and $%
\phi $ we observe that $p_{\max }^{(\lambda )}$ are 2-Lipschitz
functions,
\begin{equation*}
|p_{\max }^{(\lambda )}(\psi )-p_{\max }^{(\lambda )}(\phi )|\leq 2\Vert
\psi -\phi \Vert ,
\end{equation*}
and thus, $F_{\pi }(\psi )$ {is a }$2(d+1)$-Lipschitz function,
automatically  belonging to the Sobolev space $W^{1,\infty }(S^{2d-1})$.
The only nondifferentiable points lie in the set $\mathcal{N}$ on which at
least one basis has two equal maximal outcomes:
\begin{equation}
\mathcal{N}=\bigcup_{\lambda }\bigcup_{\kappa \neq \kappa ^{\prime }}\left\{
\psi :p_{\lambda \kappa }(\psi )=p_{\lambda \kappa ^{\prime }}(\psi
)=p_{\max }^{(\lambda )}(\psi )\right\} .  \label{NU}
\end{equation}
Each member of (\ref{NU}) is the zero set on the sphere of the real-analytic
quadratic form associated with $\langle \psi |\Pi _{\lambda \kappa }-\Pi
_{\lambda \kappa ^{\prime }}|\psi \rangle $, and thus, has Haar measure
zero. Consequently, for almost every state $|\psi \rangle $, each basis has
a unique maximizing outcome. Outside the zero-measure set (\ref{NU}) of
degenerate maxima we represent,
\begin{equation}
F_{\pi }(\psi )=\langle \psi |\hat{\Upsilon}|\psi \rangle ,\quad \hat{
\Upsilon}=\sum_{\lambda }\hat{\Pi}_{\lambda },  \label{Fpi1}
\end{equation}
where $\hat{\Pi}_{\lambda }=\hat{\Pi}_{\lambda \kappa _{\max }}$ are the
selected projectors maximizing (\ref{Fpi}).

A tangent vector $|\delta \psi \rangle $ to the unit sphere obeys $\func{Re}
\langle \delta \psi |\psi \rangle =0$, so that the first variation of (\ref{Fpi1}) is
\begin{align*}
\delta F_{\pi }(\psi )& =2\func{Re}\langle \delta \psi |\hat{\Upsilon}|\psi
\rangle \\
& =2\func{Re}\langle \delta \psi |(\hat{\Upsilon}-F_{\pi }(\psi )\hat{I}
)|\psi \rangle ,
\end{align*}
or, formally,
\begin{equation*}
\partial _{t}F_{\pi }\left( \frac{\psi +tv}{||\psi +tv||}\right) _{t=0}=2%
\func{Re}\langle v|(\hat{\Upsilon}-F_{\pi }(\psi )\hat{I})|\psi \rangle ,
\end{equation*}
where $\func{Re}\langle v|\psi \rangle =0$, since the orthogonal projection of $\hat{\Upsilon}|\psi \rangle $ onto the
tangent space is
\begin{equation*}
(I-|\psi \rangle \langle \psi |)\hat{\Upsilon}|\psi \rangle =(\hat{\Upsilon}%
-F_{\pi }(\psi )\hat{I})|\psi \rangle .
\end{equation*}
This allows us to introduce the tangent gradient
\begin{equation}
\nabla _{\mathrm{tan}}F_{\pi }=2(\hat{\Upsilon}-F_{\pi }(\psi )\hat{I})|\psi
\rangle ,  \label{gradF}
\end{equation}
{whose }norm is
\begin{equation}
\Vert \nabla _{\mathrm{tan}}F_{\pi }\Vert ^{2}=4\left( \langle \psi |\hat{
\Upsilon}^{2}|\psi \rangle -F_{\pi }^{2}(\psi )\right) \leq 4\langle \psi |
\hat{\Upsilon}^{2}|\psi \rangle .  \label{grad}
\end{equation}
The classical gradient in Eq.~(\ref{gradF}) coincides almost everywhere with
the weak gradient of the globally Lipschitz function $F_{\pi }(\psi )$.

From the mutual unbiasedness we have for $\lambda \neq \mu $,
\begin{equation*}
\left\vert \langle \psi |\hat{\Pi}_{\lambda }\hat{\Pi}_{\mu }|\psi \rangle
\right\vert =\frac{\sqrt{p_{\lambda }p_{\mu }}}{\sqrt{d}}.
\end{equation*}
Expanding $\hat{\Upsilon}^{2}=\sum_{\lambda }\hat{\Pi}_{\lambda
}+\sum_{\lambda \neq \mu }\hat{\Pi}_{\lambda }\hat{\Pi}_{\mu }$ and applying
the triangle inequality to the cross terms gives
\begin{align}
\langle \psi |\hat{\Upsilon}^{2}|\psi \rangle & \leq F_{\pi }(\psi )+\frac{1}{\sqrt{d}}\sum_{\lambda \neq \mu }\sqrt{p_{\lambda }p_{\mu }}  \label{Y2} \\
& =F_{\pi }(\psi )+\frac{1}{\sqrt{d}}\left[ \left( \sum_{\lambda }\sqrt{p_{\lambda }}\right) ^{2}-F_{\pi }(\psi )\right] .  \notag
\end{align}
Applying the Cauchy-Schwarz inequality
\begin{equation*}
\left( \sum_{\lambda }\sqrt{p_{\lambda }}\right) ^{2}\leq (d+1)\sum_{\lambda
}p_{\lambda }=(d+1)F_{\pi }(\psi ),
\end{equation*}
we arrive at
\begin{equation}
\langle \psi |\hat{\Upsilon}^{2}|\psi \rangle \leq (1+\sqrt{d})F_{\pi }(\psi
).  \label{Y2B}
\end{equation}
For normalized measure on $S^{2d-1}$, the spherical Poincar\'{e} inequality
for Lipschitz functions on $W^{1,2}(S^{2d-1})$ reads
\begin{equation}
{\func{Var}}_{S^{2d-1}}f\leq \frac{1}{2d-1}\mathbb{E}_{S^{2d-1}}\Vert \nabla
_{\mathrm{tan}}f\Vert ^{2},  \label{PI}
\end{equation}
as $F_{\pi }(\psi )\in W^{1,\infty }(S^{2d-1})\subset W^{1,2}(S^{2d-1})$.

Substituting Eqs.~(\ref{grad}) and (\ref{Y2B}) into (\ref{PI}) gives
\begin{eqnarray*}
{\func{Var}}_{\mathrm{Haar}}F_{\pi }(\psi ) &\leq &\frac{4(1+\sqrt{d})}{2d-1}
\mathbb{E}_{\mathrm{Haar}}F_{\pi } \\
&=&\frac{4(1+\sqrt{d})}{2d-1}\frac{d+1}{d}H_{d},
\end{eqnarray*}
thus, asymptotically
\begin{equation}
{\func{Var}}_{\mathrm{Haar}}M_{\pi }={\func{Var}}_{\mathrm{Haar}}F_{\pi
}(\psi )=O\!\left( \frac{\ln d}{\sqrt{d}}\right) .  \label{VarM}
\end{equation}

\section{Clifford-orbit concentration}

In this Appendix we prove the Clifford-orbit concentration result, Eq.~(\ref{VarMC}),
for uniformly sampled Clifford unitaries acting on an arbitrary
fixed $n$-qubit pure state. This result is logically distinct from the
Haar-state concentration proved in Appendix F.

Consider a stabilizer basis $\{|s_{\Gamma ,\kappa }\rangle ,\kappa \in
\mathbb{F}_{2^{n}}\}$ associated with a curve $\Gamma \in \mathcal{L}_{n}$, where $\mathcal{L}_{n}$ is the set of all Lagrangian subspaces in $V=\mathbb{F}_{2^{n}}$ $\times \mathbb{F}_{2^{n}}$, and define
\begin{equation}
p_{\kappa }(\Gamma )=\left\vert \langle s_{\Gamma ,\kappa }|\psi \rangle
\right\vert ^{2},\quad p_{\max }(\Gamma )=\max_{\kappa }p_{\kappa }(\Gamma
)\in L^{2}(\mathcal{L}_{n}),  \label{Phip}
\end{equation}
the maximum probability of {detecting }a given state $|\psi \rangle $ in
this basis.

The Pauli subgroup of the Clifford group only permutes the eigenvalue labels
inside each stabilizer basis and does not change the maximum. Let us denote
\begin{equation}
X_{\lambda }(C) =p_{\max }(S_{C}^{-1}\Gamma ^{\lambda }),~~\mathbb{E}
_{C}X_{\lambda }(C)=\mathbb{E}_{\Gamma }p_{\max }(\Gamma ),  \label{XC}
\end{equation}
\begin{eqnarray}
\func{Var}_{C}M_{\pi } &=&\sum_{\lambda }\func{Var}_{C}X_{\lambda
}(C)\nonumber\\
&&+\sum_{\lambda \neq \mu }\func{Cov}_{C}\left( X_{\lambda }(C),X_{\mu
}(C)\right) ,  \label{VC}
\end{eqnarray}
where $\Gamma ^{\lambda }$ is the Lagrangian subspace defining the $\lambda $-th basis of $\pi $, and $S_{C}\in \func{Sp}(2n,2)$ is the symplectic action
induced by $\hat{C}$. Random Clifford transformations induce a uniform
sampling of $S_{C}$ in $\func{Sp}(2n,2)$. Since $\func{Sp}(2n,2)$ acts
transitively on $\mathcal{L}_{n}$, for every fixed $\lambda $, the random
Lagrangian subspace $\Gamma _{C}^{\lambda }=S_{C}^{-1}\Gamma ^{\lambda }$ is
uniform on $\mathcal{L}_{n}$. In addition, curves forming a phase-space
partition are transverse,
\begin{equation}
\Gamma ^{\lambda }\cap \Gamma ^{\mu }=\{0\},\quad \lambda \neq \mu ,
\label{noInsec}
\end{equation}
and as the symplectic group acts transitively on ordered transverse pairs,
every pair $(\Gamma _{C}^{\lambda },\Gamma _{C}^{\mu })$ is uniformly
distributed over all ordered pairs $(\Gamma ,\Gamma ^{\prime })$ satisfying
(\ref{noInsec}).

For a fixed $\Gamma $ one has a simple inequality
\begin{equation}
p_{\max }^{2}(\Gamma )\leq \left( \sum_{\kappa =1}^{d}p_{\kappa }^{3}(\Gamma
)\right) ^{2/3}.  \label{p3}
\end{equation}
Averaging (\ref{p3}) over uniformly distributed stabilizer bases $\Gamma $,
and using the concavity of $x^{2/3}$ gives the following relation for the
expectation values
\begin{equation}
\mathbb{E}_{\Gamma }p_{\max }^{2}(\Gamma )\leq \left[ \mathbb{E}_{\Gamma
}\sum_{\kappa =1}^{d}p_{\kappa }^{3}(\Gamma )\right] ^{2/3},  \label{Jensen}
\end{equation}
where a uniformly random pair ($\Gamma ,\kappa $) defines a uniformly
random stabilizer state. Using the 3-design property of stabilizer states
\cite{zhu17,Webb}:
\begin{equation*}
\mathbb{E}_{s\,\mathrm{stab}}\left\vert \langle s|\psi \rangle \right\vert
^{6}=\frac{6}{d(d+1)(d+2)},
\end{equation*}
we obtain
\begin{equation}
\mathbb{E}_{\Gamma }\sum_{\kappa =1}^{d}p_{\kappa }(\Gamma )^{3}=d\,\mathbb{E
}_{s\,\mathrm{stab}}|\langle s|\psi \rangle |^{6}=\frac{6}{(d+1)(d+2)},
\end{equation}
so that,
\begin{equation}
\mathbb{E}_{\Gamma }p_{\max }^{2}(\Gamma )\leq a_{d},\qquad a_{d}=\left[
\frac{6}{(d+1)(d+2)}\right] ^{2/3},  \label{ad}
\end{equation}
and in particular,
\begin{eqnarray}
\sum_{\lambda }\func{Var}_{C}X_{\lambda }(C)&=&(d+1)\func{Var}_{\Gamma
}p_{\max }(\Gamma )\nonumber\\
&\leq& (d+1)a_{d}=O(d^{-1/3}).  \label{varG}
\end{eqnarray}
To estimate the covariance in (\ref{VC}) for $\lambda \neq \mu $, take into
account that $(\Gamma ,\Gamma ^{\prime })$ is uniformly distributed over
ordered transverse pairs of Lagrangian subspaces,
\begin{equation*}
\func{Cov}_{C}\left( X_{\lambda }(C),X_{\mu }(C)\right) =\mathbb{E}_{\Gamma
,\Gamma ^{\prime }}[p_{\max }^{c}(\Gamma )p_{\max }^{c}(\Gamma ^{\prime })],
\end{equation*}
where the centered probabilities are%
\begin{equation}
p_{\max }^{c}(\Gamma )=p_{\max }(\Gamma )-\mathbb{E}_{\Gamma }[p_{\max
}(\Gamma )],  \label{pC}
\end{equation}
we use the normalized averaging operator of a function $g$ defined on a
curve $\Gamma $ over all transverse curves \cite{Eisfeld}
\begin{equation}
(Tg)(\Gamma )=\frac{1}{k_{n}}\sum_{\substack{ \Gamma ^{\prime }\in \mathcal{L
}_{n}  \\ \Gamma ^{\prime }\cap \Gamma =\{0\}}}g(\Gamma ^{\prime }),\quad T
\mathbf{1}=\mathbf{1}  \label{Tav}
\end{equation}
where $k_{n}=2^{n(n+1)/2}$ is the number of curves transverse to every $\Gamma $. Equivalently, if $\Gamma ^{\prime }$ is chosen uniformly among the
curves transverse to a fixed $\Gamma $, then
\begin{equation}
(Tg)(\Gamma )=\mathbb{E}\!\left[ g(\Gamma ^{\prime })\,|\,\Gamma ^{\prime
}\cap \Gamma =\{0\},\Gamma \right] ,  \label{Tcond}
\end{equation}
so that
\begin{equation}
\mathbb{E}_{\Gamma }(Tg)(\Gamma )=\mathbb{E}_{\Gamma }g(\Gamma ).  \label{ET}
\end{equation}
Real functions on $\mathcal{L}_{n}$ are {equipped }with the inner product
\begin{equation}
\langle g,h\rangle =\mathbb{E}_{\Gamma }[g(\Gamma )h(\Gamma )]=\frac{1}{|
\mathcal{L}_{n}|}\sum_{\Gamma \in \mathcal{L}_{n}}g(\Gamma )h(\Gamma ),
\label{TIP}
\end{equation}
under which the averaging operator (\ref{Tav}) is symmetric and self-adjoint
\cite{Eisfeld}, $\langle g,Th\rangle =\langle Tg,h\rangle $. The operator $T$
quantifies the amount of information about a given $\Gamma $ that survives
after averaging under the transversality constraint.
In particular, if $\Gamma ^{\prime }$ were sampled independently of $\Gamma $, then the
conditional average $\mathbb{E}_{\Gamma ^{\prime }}g(\Gamma ^{\prime })$
in Eq.~(\ref{Tcond}) would be the constant. Thus, only the variable part of $Tg$
 generates a covariance between observables evaluated at $\Gamma $ and $\Gamma ^{\prime }$. The spectrum of $T$, given by \cite{Eisfeld}
\begin{equation*}
t_{r}=(-1)^{r}2^{-r(n-r+1)},\qquad r=0,1,\ldots ,n,
\end{equation*}
provides the bound on this residual dependence, where $t_{0}=1$ corresponds
to the uniform average of the function, in accordance with \ref{Tav}. The
remaining spectral values satisfy the restrictions
\begin{equation*}
|t_{r}|\leq 2^{-n}=d^{-1},\qquad r=1,\ldots ,n.
\end{equation*}
Consequently, the operator norm of $T$ restricted to the zero-mean subspace $\mathbf{1}^{\perp }=\{g\in L^{2}(\mathcal{L}_{n}):\mathbb{E}_{\Gamma
}[g(\Gamma )]=0\}$ is bounded by:
\begin{equation}
\Vert T\Vert _{\mathbf{1}^{\perp }}=\max_{1\leq r\leq n}|t_{r}|=d^{-1}.
\label{Tnorm}
\end{equation}
The centered probability (\ref{pC}) belongs, by construction, to $\mathbf{1}^{\perp }$: $p_{\max }^{c}(\Gamma )\in \mathbf{1}^{\perp }$. For a
uniformly distributed ordered transverse pair $(\Gamma ,\Gamma ^{\prime })$,
conditioning on $\Gamma $ expectation values are
\begin{equation*}
\mathbb{E}_{\Gamma ,\Gamma ^{\prime }}\!\left[ p_{\max }^{c}(\Gamma )p_{\max
}^{c}(\Gamma ^{\prime })\right] =\mathbb{E}_{\Gamma }\!\left[ p_{\max
}^{c}(\Gamma )\mathbb{E}\!\left[ p_{\max }^{c}(\Gamma ^{\prime })\mid \Gamma
\right] \right]
\end{equation*}
\begin{equation*}
=\mathbb{E}_{\Gamma }\!\left[ p_{\max }^{c}(\Gamma )(Tp_{\max }^{c})(\Gamma )
\right] =\langle p_{\max }^{c}(\Gamma ),Tp_{\max }^{c}(\Gamma )\rangle .
\end{equation*}
Applying the Cauchy-Schwarz inequality along with the spectral norm bound (\ref{Tnorm}) we arrive at
\begin{align*}
{|\func{Cov}_{C}(X_{\lambda },X_{\mu })|}& {\leq \Vert p_{\max }^{c}(\Gamma
)\Vert _{2}\,\Vert Tp_{\max }^{c}(\Gamma )\Vert _{2}} \\
& {\leq \frac{1}{d}\Vert p_{\max }^{c}(\Gamma )\Vert _{2}^{2}=\frac{1}{d}
\func{Var}_{\Gamma }p_{\max }(\Gamma )},
\end{align*}
 leading to the geometric covariance estimate:
\begin{equation}
|\func{Cov}_{C}(X_{\lambda },X_{\mu })|\leq \frac{1}{d}\func{Var}
_{C}X_{\lambda }\leq \frac{a_{d}}{d},\quad \lambda \neq \mu .  \label{CovT}
\end{equation}
Finally, we obtain from (\ref{VC}), (\ref{varG}) and (\ref{CovT})
\begin{eqnarray*}
\func{Var}_{C}M_{\pi }\leq (d+1)a_{d}+d\left( d+1\right) \frac{a_{d}}{d}
&=&2(d+1)a_{d}\\
&=&O(d^{-1/3}),
\end{eqnarray*}
which vanishes as {$O(2^{-n/3})$}. By Chebyshev's inequality, for every
fixed $\varepsilon >0$, we have
\begin{equation*}
\Pr_{C}\!\left( \left\vert M_{\pi }(\hat{C}|\psi \rangle )-\mathbb{E}
_{C}M_{\pi }(\hat{C}|\psi \rangle )\right\vert >\varepsilon \right) \leq
\frac{2(d+1)a_{d}}{\varepsilon ^{2}}\longrightarrow 0.
\end{equation*}
In particular, if $\hat{C}$ and $\hat{C}^{\prime }$ are sampled
independently and uniformly from the Clifford group, then
\begin{equation*}
M_{\pi }(\hat{C}|\psi \rangle )-M_{\pi }(\hat{C}^{\prime }|\psi \rangle )
\overset{\Pr }{\rightarrow }0,
\end{equation*}
that is, the witness becomes statistically insensitive to the random
Clifford frame.

\section{Sparse-state linear bounds}

In this Appendix we prove the linear-order behavior (\ref{Mn}) for a
class of sparse states.

Let us consider states with affinely independent supports $\mathbb{S}\subset
\mathbb{F}_{2}^{n}$, such that for any chosen $\tau _{0}\in \mathbb{S}$
the vectors $\{\tau -\tau _{0}:\tau \in \mathbb{S}\backslash \{\tau _{0}\}\}$
are linearly independent over {{$\mathbb{F}_{2}$. This automatically
restricts the support cardinality to $|\mathbb{S}|\leq n+1$}} . We
additionally impose the condition on the size of the effective support
\begin{equation}
{S_{\mathrm{eff}}(\psi )=\left( \sum_{\tau \in \mathbb{S}}|c_{\tau
}|^{4}\right) ^{-1}=\Theta (n).}  \label{Seff}
\end{equation}
\textit{Proposition (sparse-state bounds).} For an $n$-qubit pure state
with affinely independent computational support, every Clifford unitary $\hat{C}$ the following inequality is satisfied
\begin{equation}
\frac{2}{\pi ^{2}}\left( 1+d^{-1}\right) ^{2}r^{2}(\psi )-1\leq M_{\pi }(\hat{C}|\psi \rangle )\leq r^{2}(\psi ),  \label{twosided}
\end{equation}
where $r(\psi )=\sum_{\tau \in \mathbb{S}}|c_{\tau }|$.

The relations $S_{\mathrm{eff}}(\psi )\leq r^{2}(\psi )\leq |\mathbb{S}|$
and (\ref{Seff}) imply $r^{2}(\psi )=\Theta (n)$, therefore Eq. (\ref{twosided}) leads to Eq. (\ref{Mn}).

\textit{Proof.} The upper bound follows from Eqs.~(\ref{MD}) and (\ref{Dbound}). Since $\mathcal{D}(\hat{\rho})$ is Clifford invariant,
\begin{equation}
M_{\pi }(\hat{C}|\psi \rangle )\leq \mathcal{D}(\hat{C}|\psi \rangle )=
\mathcal{D}(|\psi \rangle )\leq r^{2}(\psi ).  \label{upper}
\end{equation}
To obtain a lower bound let us consider one particular stabilizer $\lambda $-basis of the Clifford-rotated complete MUB.
The elements of this basis $\{|\psi _{\kappa }^{\lambda }(C)\rangle \}$ are supported on disjoint affine
subsets $H_{\lambda }^{(m)}$, $m=1,..,2^{r_{\lambda }}$ of the computational
basis, each of cardinality $|H_{\lambda }|=2^{-r_{\lambda }}d$, where $r_{\lambda }=\dim _{\mathbb{F}_{2}}(\Gamma _{\lambda }\cap \Gamma _{Z})$ is
the number of diagonal operators of the stabilizer set. Thus, $2^{n-r_{\lambda }}$ basis vectors $\{|\psi _{\kappa ,H}^{\lambda }\rangle \}$
share the same support.

Since $\mathbb{S}$ is affinely independent, every subset $H_{\lambda
}^{(m)}\cap \mathbb{S}$ is affinely independent, and the characters labeling
the stabilizer vectors supported on $H$ realize all relative sign patterns.
Thus, in each $H_{\lambda }^{(m)}$-sector one has
\begin{eqnarray}
&&\max_{\kappa }|\langle \psi _{\kappa ,H}^{\lambda }|\psi \rangle |\nonumber\\
&=&|H_{\lambda }^{(m)}|^{-1/2}\max_{\varepsilon _{\tau }=\pm 1}\left\vert
\sum_{\tau \in H_{\lambda }^{(m)}\cap \mathbb{S}}\varepsilon _{\tau }\phi
^{\lambda \ast }(\tau )c_{\tau }\right\vert  \label{signmax} \\
&\geq &\frac{2}{\pi \sqrt{|H_{\lambda }^{(m)}|}}\sum_{\tau \in H_{\lambda
}^{(m)}\cap \mathbb{S}}|c_{\tau }|,
\end{eqnarray}
where $2/\pi $ factor comes from averaging $\varepsilon _{\tau }(\theta )=
\func{sgn}\func{Re}[e^{-i\theta }\phi ^{\lambda \ast }(\tau )c_{\tau }]$
over $0\leq \theta <\pi $, where $\phi ^{\lambda }(\tau )$ is the cocycle
phase. The above equation implies
\begin{equation}
p_{\max }^{(\lambda )}\geq \frac{4}{\pi ^{2}}\frac{l^{2}(H_{\lambda }^{(m)})
}{|H_{\lambda }^{(m)}|},\quad l(H_{\lambda }^{(m)})=\sum_{\tau \in
H_{\lambda }^{(m)}\cap \mathbb{S}}|c_{\tau }|.  \label{fixedH}
\end{equation}
Since the sum over $2^{r_{\lambda }}$ supports gives $\sum_{m}l(H_{\lambda
}^{(m)})=r(\psi )$, at least one support satisfies $l(H_{\lambda
}^{(m)})\geq 2^{-r_{\lambda }}r(\psi )$, so that
\begin{equation}
p_{\max }^{(\lambda )}\geq \frac{4}{\pi ^{2}}\frac{r^{2}(\psi )}{d}
\,2^{-r_{\lambda }}.  \label{pmaxlower}
\end{equation}
It follows from Eq.(\ref{sumH-1}) that
\begin{equation*}
\sum_{\lambda =1}^{d+1}2^{r_{\lambda }}=2d,
\end{equation*}
thus, the Cauchy-Schwarz inequality gives
\begin{equation}
\sum_{\lambda =1}^{d+1}2^{-r_{\lambda }}\geq \frac{(d+1)^{2}}{\sum_{\lambda
}2^{r_{\lambda }}}=\frac{(d+1)^{2}}{2d}.  \label{E:inverseq}
\end{equation}
Finally, summing Eq.~(\ref{pmaxlower}) we arrive at the left hand side
inequality (\ref{twosided}).

\section{Heuristic active-sector scrambling approximation}

Appendix H proves linear-order behavior for affinely independent sparse
supports with extensive effective supports. Numerically, however, the same
linear behavior is observed for considerably broader families, including
states with exponentially large computational support. The purpose of this
Appendix is to give a heuristic, physical explanation of that observation.

A general n-qubit state may have a stabilizer sector, which trivially
contributes to the value of measure (\ref{M}). Thus, we consider a class of $n$-qubit states $|\psi _{n}\rangle $
with a stabilizer subgroup $\mathcal{S}_{P}(\psi _{n})$ containing $s$ independent generators, and therefore, under
Clifford transformations, reducible to the decoupled form
\begin{equation*}
|\psi _{n}\rangle =\hat{C}_{0}\left( |0\rangle ^{\otimes s}\otimes |\chi
_{t}\rangle \right) ,\quad \hat{C}_{0}\in \mathcal{C},
\end{equation*}
with a large stabilizer nullity
\begin{equation*}
t(\psi _{n})=n-s=an+o(n),\quad 0<a\leq 1.
\end{equation*}
In practice, to ensure robustness against small non-Clifford perturbations,
the approximate nullity \cite{beverland} is used.

We focus on nonsparse families of states with large active $t$-qubit sectors
$|\chi _{t}\rangle $, delocalized over exponentially many computational
configurations:
\begin{equation*}
-\ln \sum_{\kappa \in \mathrm{Supp}(\chi _{t})}|\chi _{t}(\kappa )|^{4}\geq
ct,\quad \max_{\kappa }|\chi _{t}(\kappa )|^{2}\leq 2^{-\eta t+o(t)},
\end{equation*}
with $\chi _{t}(\kappa )=\langle \kappa |\chi _{t}\rangle $, and $c,\eta >0$. Thus, the number of active configurations is at least exponentially large,
$|\mathrm{Supp}(\chi _{t})|\geq 2^{\eta t-o(t)}$.

A Clifford transformation $\hat{C}$ maps the Lagrangian subspaces $\{\Gamma^{\lambda }\}$, labeling the original MUBs, into $\{\Gamma _{C}^{\lambda }\}$
, and the intersection dimension $j_{\lambda }(C)=\dim _{\mathbb{F}_{2}}(\Gamma _{C}^{\lambda }\cap S_{\psi })$ counts redundant measurements
already fixed by the stabilizers of the state. Since the state is stabilized
by $S_{\psi }$, and thus, $\langle \psi _{n}|\hat{D}_{u}|\psi _{n}\rangle =0$, $u\notin S_{\psi }^{\perp }$, the symplectic pairing from $\Gamma
_{C}^{\lambda }$ to the dual of $S_{\psi }$ (whose kernel is $\Gamma
_{C}^{\lambda }\cap S_{\psi }^{\perp }$) has rank $s-j_{\lambda }$.
Consequently, the operators that produce nontrivial measurement outcomes are
labelled by $\Gamma _{C}^{\lambda }\cap S_{\psi }^{\perp }$ with
\begin{equation}
\dim \left( \Gamma _{C}^{\lambda }\cap S_{\psi }^{\perp }\right)
=n-(s-j_{\lambda })=t+j_{\lambda },
\end{equation}
by the rank-nullity theorem. Quotienting out the $j_{\lambda }(C)$ redundant
operators, $\left( \Gamma _{C}^{\lambda }\cap S_{\psi }^{\perp }\right)/(\Gamma _{C}^{\lambda }\cap S_{\psi })\simeq \mathbb{F}_{2}^{t}$ leaves $t$
independent commuting displacements acting on the active logical qubits with
rank-one projection operators $\{\hat{\Pi}_{b}^{(\lambda ,C)}=|\varphi
_{b}^{(\lambda ,C)}\rangle \langle \varphi _{b}^{(\lambda
,C)}|\}_{b=1}^{2^{t}}$, and yielding the outcome
probabilities
\begin{eqnarray}
q_{b}^{(\lambda ,C)}(\chi _{t})&=&\langle \chi _{t}|\hat{\Pi}_{b}^{(\lambda
,C)}|\chi _{t}\rangle\nonumber\\
&=&\frac{1}{|H_{\lambda ,C}|}|\sum_{\kappa \in
H_{\lambda ,C}}\chi _{t}(\kappa )\Phi _{b}^{\lambda ,C}\left( \kappa \right)
|^{2},  \label{qprob}
\end{eqnarray}
\begin{equation*}
\sum_{b=1}^{2^{t}}q_{b}^{(\lambda ,C)}(\chi _{t})=1,
\end{equation*}
where the stabilizer vector $|\varphi _{b}^{(\lambda ,C)}\rangle $ is
supported on $H_{\lambda ,C}\subseteq \mathbb{F}_{2}^{t}$, and $\Phi
_{b}^{\lambda ,C}\left( \kappa \right) $ are the corresponding cocycle
phases.

Since $s-j_{\lambda }$ measurement directions, that do not overlap with the
state's stabilizers $S_{\psi }$, produce unbiased outcomes, contributing by
a factor $2^{-[s-j_{\lambda }(C)]}$, the maximum probabilities in each
rotated basis are
\begin{equation}
p_{\max }^{(\lambda )}(\hat{C}|\psi _{n}\rangle )=2^{-[s-j_{\lambda
}(C)]}q_{\max }^{(\lambda ,C)}(\chi _{t}),  \label{pqmax}
\end{equation}
where $q_{\max }^{(\lambda ,C)}(\chi _{t})$ is the largest probability in
the active $t$-qubit sector; in particular, $p_{\max }^{(\lambda
)}=2^{-(n-t)}q_{\max }^{(\lambda ,C)}$ if $\Gamma _{C}^{\lambda }\cap
S_{\psi }=\{0\}$.

Note that since every nonzero element of $S_{\psi }$ belongs to exactly one
of $d+1$ subspaces $\Gamma _{C}^{\lambda }$ of the full partition, and the
number of non-zero elements in each $\Gamma _{C}^{\lambda }\cap S_{\psi }$
is $2^{j_{\lambda }(C)}-1$, while its total number is $2^{s}-1$, one has
\begin{equation*}
\sum_{\lambda =1}^{d+1}\left( 2^{j_{\lambda }(C)}-1\right) =2^{s}-1,
\end{equation*}
and thus,
\begin{equation}
\sum_{\lambda =1}^{d+1}2^{j_{\lambda }(C)}=d+2^{s}.  \label{jsum}
\end{equation}

We now proceed with the physical approximation underlying the extreme-value
estimate: phase scrambling in the active sector. Expanding Eq.~(\ref{qprob}), the off-diagonal interference contribution in a fixed channel $b$ is
\begin{eqnarray}
&&\sum_{\kappa \neq \kappa ^{\prime }}\chi _{t}(\kappa )\chi _{t}^{\ast
}(\kappa ^{\prime })\Phi _{b}^{\lambda ,C}(\kappa )\Phi _{b}^{\lambda
,C}(\kappa ^{\prime })^{\ast }\nonumber\\
&&=\sum_{\kappa \neq \kappa ^{\prime }}|\chi
_{t}(\kappa )|\,|\chi _{t}(\kappa ^{\prime })|e^{i\delta _{b}^{\lambda
,C}(\kappa ,\kappa ^{\prime })},  \label{offdiagF}
\end{eqnarray}
where the microscopic phase mismatch $\delta _{b}^{\lambda ,C}(\kappa
,\kappa ^{\prime })$ combines the phases of the active-sector amplitudes
with the Clifford cocycle phases. For a generic non-Clifford active sector
these mismatches are not systematically phase locked to the cocycle phases.
We first coarse grain microscopic terms contributing to active-sector
overlaps $\langle \varphi _{b}^{(\lambda ,C)}|\chi _{t}\rangle $ over an
extensive number of active degrees of freedom, by partitioning them into
$m_{t}=\alpha _{\psi }t+o(t)$ effective blocks of comparable amplitudes
$w_{l} $ within which the phases remain strongly correlated, but the phases
$\{\Delta _{\ell }^{(b)}:\ell =1,...,m_{t}\}$ of different blocks are weakly
correlated, with no preferred direction, $\mathbb{E}(e^{i\Delta _{\ell}^{(b)}})=0$:
\begin{equation}
2^{t/2}\langle \varphi _{b}^{(\lambda ,C)}|\chi _{t}\rangle \approx
\sum_{\ell =1}^{m_{t}}w_{l}e^{i\Delta _{\ell }^{(b)}}\approx
m_{t}^{1/2}A_{t}^{(b)},  \label{qA}
\end{equation}
where $w_{l}\sim m_{t}^{-1/2}$, and thus, the empirical averages
\begin{equation}
A_{t}^{(b)}=m_{t}^{-1}\sum_{\ell =1}^{m_{t}}e^{i\Delta _{\ell }^{(b)}},
\label{phasorF}
\end{equation}
are typically small, $\mathbb{E}|A_{t}^{(b)}|^{2}=O(m_{t}^{-1})$, expressing
the usual random-phase cancellation.

Approximating the coarse-grained phase blocks $\Delta _{\ell }^{(b)}$ as
independent at leading exponential order, Cram\'{e}r's theorem \cite{Varadhan1984, Touchette2009}
applied to random two-dimensional vectors $(\cos \Delta _{\ell },\sin \Delta _{\ell })$, gives
\begin{equation}
\Pr \{A_{t}^{(b)}\simeq z\}=\exp [-m_{t}J_{\psi }(z)+o(t)],\quad z\in
\mathbb{C},  \label{At}
\end{equation}
where $J_{\psi }(z)\geq 0$ is a phase-alignment rate function.

However, the phases of different outcome channels can still be partially
correlated, so, the corresponding outcome probabilities $q_{b}^{(\lambda,C)} $
are not mutually independent. In other words, the corresponding
cross-term interference patterns are not completely independent. These
correlations reduce the effective number of statistically distinct phase
patterns to
\begin{equation}
N_{\mathrm{eff}}=\exp [\eta _{\psi }t+o(t)],\qquad 0<\eta _{\psi }\leq \ln 2.
\label{NeffF}
\end{equation}
The limiting value $\eta _{\psi }\rightarrow 0$ corresponds to a
subexponential number of patterns, with $N_{\mathrm{eff}}=O(1)$ in the
Clifford-like synchronized limit, whereas $\eta _{\psi }=\ln 2$ corresponds
to the Haar-like limit in which all $2^{t}$ measurement channels are
asymptotically distinct at the exponential scale.

It is convenient to introduce the rescaled probability
\begin{equation}
Y_{b}^{(\lambda ,C)}=2^{t}q_{b}^{(\lambda ,C)}(\chi _{t}),
\end{equation}
normalized to
\begin{equation}
\mathbb{E}_{b}[Y_{b}^{(\lambda ,C)}]\equiv \frac{1}{2^{t}}
\sum_{b=1}^{2^{t}}Y_{b}^{(\lambda ,C)}=1.
\end{equation}
For a flat distribution, $q_{b}^{(\lambda ,C)}(\chi _{t})=2^{-t}$,
$Y_{b}^{(\lambda ,C)}=1$ for every $b$, whereas $Y_{b}^{(\lambda ,C)}=O(t)$
represents a large peak above the flat incoherent scale and requires an
atypical coherent fluctuation $|A_{t}^{(b)}|=O(1)$, whose probability is
exponentially small according to (\ref{At}). It follows from (\ref{phasorF})
and (\ref{qA}) that
\begin{equation*}
\Pr_{C,b}\{Y_{b}^{(\lambda ,C)}>xt\}\simeq
\Pr_{C,b}\{|A_{t}^{(b)}|^{2}>x/\alpha _{\psi }\},
\end{equation*}
as $Y_{b}^{(\lambda ,C)}\approx m_{t}|A_{t}^{(b)}|^{2}$, where $\Pr_{C,b}$
denotes joint sampling of a random Clifford frame and an outcome channel.
The dominant contribution comes from the least costly alignment, i.e.,
$I_{\psi }(x)=\alpha _{\psi }\inf_{|z|^{2}\geq x/\alpha _{\psi }}J_{\psi }(z)$, giving for a single outcome channel
\begin{equation}
\Pr_{C,b}\{Y_{b}^{(\lambda ,C)}>xt\}=\exp [-tI_{\psi }(x)+o(t)],
\label{PYgeneral}
\end{equation}
where no mutual independence of channels is assumed; their residual
correlations are encoded in $N_{\mathrm{eff}}$.

The maximum intensity $Y_{\max }=\max_{b}Y_{b}$ is determined by the
threshold where the expected number of effective channels achieving such a
constructive fluctuation is of order unity:
\begin{equation}
N_{\mathrm{eff}}\,\Pr \{Y_{b}^{(\lambda ,C)}>x_{\ast }t\}\asymp \exp
\{t[\eta _{\psi }-I_{\psi }(x_{\ast })]+o(t)\}\asymp 1,
\end{equation}
giving $I_{\psi }(x_{\ast })=\eta _{\psi }$, and
\begin{equation}
Y_{\max }^{(\lambda ,C)}=x_{\ast }t+o_{\Pr }(t)=\Theta (t),  \label{YmaxF}
\end{equation}
provided there is a finite solution $x_{\ast }>0$. Thus, the maximum
probabilities in the active sectors are
\begin{equation}
q_{\max }^{(\lambda ,C)}(\chi _{t})=2^{-t}Y_{\max }^{(\lambda ,C)}=\Theta
\left( \frac{t}{2^{t}}\right) .  \label{qmaxF}
\end{equation}
Using (\ref{pqmax}), the witness has the exact active-sector decomposition
\begin{equation*}
M_{\pi }(\hat{C}|\psi _{n}\rangle )+1=2^{t-n}\sum_{\lambda
=1}^{d+1}2^{j_{\lambda }(C)}q_{\max }^{(\lambda ,C)}(\chi _{t}).
\end{equation*}
Assuming that the order-one coefficient implicit in (\ref{qmaxF}) remains
uniformly of order unity for the bases carrying non-negligible weight in the
partition sum, Eqs. (\ref{qmaxF}) and (\ref{jsum}) give
\begin{eqnarray*}
M_{\pi }(\hat{C}|\psi _{n}\rangle )+1 &=&\Theta \left( 2^{t-n}\frac{t}{2^{t}}
(d+2^{s})\right) \\
&=&\Theta \left( t(1+2^{-t})\right) ,
\end{eqnarray*}
that is,
\begin{equation}
M_{\pi }(\hat{C}|\psi _{n}\rangle )=\Theta (t),  \label{MFactive}
\end{equation}%
and an extensive robust active sector $t=an+o(n)$ heuristically yields
$M_{\pi }(\hat{C}|\psi _{n}\rangle )=\Theta (n)$.

Two examples illustrate the intended regime. First,
\begin{equation*}
|T\rangle ^{\otimes \rho n}\otimes |+\rangle ^{\otimes (1-\rho )n}
\end{equation*}%
has $t=\rho n$ active non-Clifford qubits, so the argument predicts $M_{\pi
}=\Theta (\rho n)$. Second, the half-filled Dicke state has exponentially
large support
\begin{eqnarray*}
|\mathrm{Supp}(D_{n/2}^{n})|&=&\binom{n}{n/2},\\
 \log _{2}\binom{n}{n/2}&=&n-%
\frac{1}{2}\log _{2}n+O(1),
\end{eqnarray*}%
and for even $n>2$, has only two independent stabilizers: $X^{\otimes n}$
and $Z^{\otimes n}$ , thus $s=2$ and $t=n-2=\Theta (n)$, belonging to the
delocalized extensive-support regime. In general, the family of Dicke states
$\mathrm{|}D_{k}^{n\gg 1}\rangle $, $k/n=\varsigma $, $0<\varsigma <1$, lies
in the extensive active sector, consistently with the linear behavior
observed numerically.

\begin{figure}[tbp]
\includegraphics[width=\linewidth]{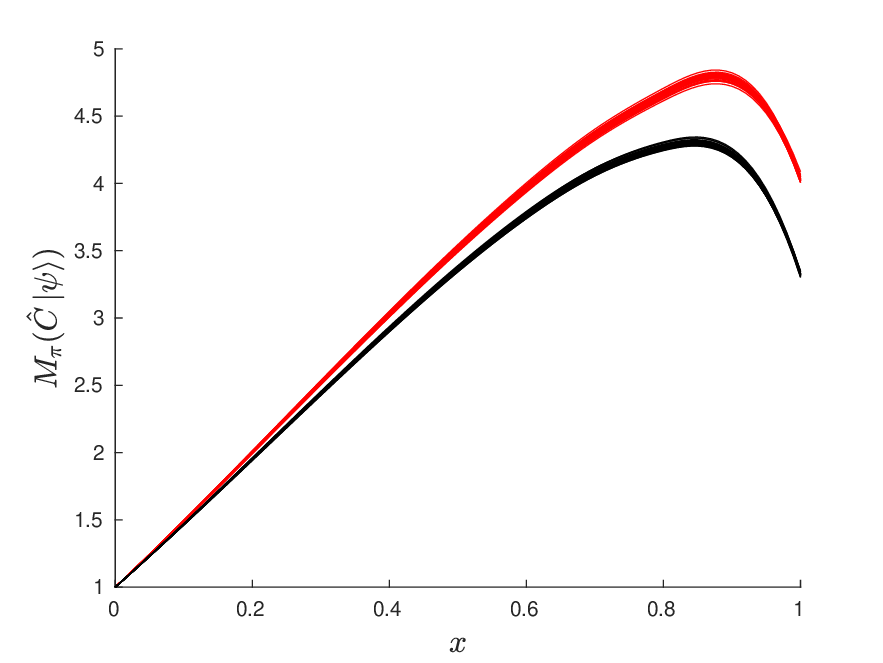}
\caption{The witness (\ref{M}) for the states $\sqrt{1-x^{2}}|CS\rangle +x|W\rangle $ (red curves), and $\sqrt{1-x^{2}}|CS\rangle
+x|k=n/2,n\rangle $ (black curves), under the action of $30$ random Clifford
transformations, $n=10$ \cite{data}.}
\label{fig:fig12}
\end{figure}

We numerically observe that near stabilizer states, the witness (\ref{M}) is
also approximately Clifford-invariant, as well as for states ranging from
polynomial to exponentially large computational support. In Fig. \ref{fig:fig12},
 we plot the witness (\ref{M}) for Clifford-transformed
superpositions of the cluster state (\ref{psimu}) with the W-state with
support size $|\mathbb{S}|=n$ (red curves), and with the Dicke state
$|k=n/2,n\rangle $, whose support is $\sim 2^{n}/\sqrt{n}$ (black curves). In
both cases, the witness remains reasonably stable under Clifford
transformations, even for a moderate number of qubits.

\end{document}